\documentclass[12pt]{article} 
\usepackage{amsmath,amsthm}
\usepackage{eufrak}

\DeclareMathAlphabet{\bb}{U}{msb}{m}{n} \gdef\C{\bb C} \gdef\dZ{\bb
Z}   \gdef\dS{\bb S} \gdef\R{\bb R}
\gdef\K{\bb K} \gdef\BH{\bb H} \gdef\F{\bb F} 

\DeclareMathOperator{\End}{End} \DeclareMathOperator{\spin}{{\bf
Spin}} \DeclareMathOperator{\pin}{{\bf Pin}}

\DeclareMathOperator{\Id}{Id} 
\DeclareMathOperator{\Aut}{PT} \DeclareMathOperator{\sAut}{{\sf
PT}} \DeclareMathOperator{\sExt}{{\sf CPT}}
\DeclareMathOperator{\Ker}{Ker}

\DeclareMathOperator{\Ext}{CPT} \DeclareMathOperator{\Mat}{Mat}

\newcommand{\s}{\!}

\newcommand{\bi}{{\bf i}}
\newcommand{\cA}{\mathcal{A}}

\newcommand{\cE}{\mathcal{E}}

\newcommand{\sA}{{\sf A}}
\newcommand{\sB}{{\sf B}}
\newcommand{\sI}{{\sf I}}

\newcommand{\sW}{{\sf W}}
\newcommand{\sE}{{\sf E}}
\newcommand{\sC}{{\sf C}}
\newcommand{\sF}{{\sf F}}

\newcommand{\sS}{{\sf S}}

\newcommand{\sK}{{\sf K}}

\newcommand{\bj}{{\bf j}}
\newcommand{\bk}{{\bf k}}

\newcommand{\bx}{{\bf x}}

\newcommand{\bZ}{{\bf Z}}

\newcommand{\fM}{\mathfrak{M}}

\newcommand{\fC}{\mathfrak{C}}
\newcommand{\fR}{\mathfrak{R}}
\newcommand{\fH}{\mathfrak{H}}

\newcommand{\cl}{C\kern -0.2em \ell}

\newcommand{\p}{\prime}
\newcommand{\e}{\mbox{\bf e}}

\newtheorem{theorem}{Theorem}
\begin{document}
\title{$CPT$ Groups of Spinor Fields in de Sitter and Anti-de Sitter Spaces}
\author{V.~V. Varlamov
\thanks{
Siberian State Industrial University,
Kirova 42, Novokuznetsk 654007, Russia}}
\date{}
\maketitle
\begin{abstract}
$CPT$ groups for spinor fields in de Sitter and anti-de Sitter
spaces are defined in the framework of automorphism groups of
Clifford algebras. It is shown that de Sitter spaces with mutually
opposite signatures correspond to Clifford algebras with different
algebraic structure that induces an essential difference of $CPT$
groups associated with these spaces. $CPT$ groups for charged
particles are considered with respect to phase factors on the
various spinor spaces related with real subalgebras of the simple
Clifford algebra over the complex field (Dirac algebra). It is shown
that $CPT$ groups for neutral particles which admit
particle-antiparticle interchange and $CPT$ groups for truly neutral
particles are described within semisimple Clifford algebras with
quaternionic and real division rings, respectively. A difference
between bosonic and fermionic $CPT$ groups is discussed.
\end{abstract}
\maketitle

\section{Introduction}
As is known, de Sitter and anti-de Sitter spaces of different
dimensions have been extensively studied during the past two decades
mainly with the purpose of understanding the correspondence between
supergravity in a five-dimensional anti-de Sitter space and $N=4$
supersymmetric field theory in four dimensions. On the other hand,
quantum field theory on the de Sitter spacetime by itself at present
time is a rapid developing area in theoretical and mathematical
physics \cite{Mie77}--\cite{BGMT01}. One of the first problem, which
naturally arising in this context, is an investigation of $CPT$
groups for spinor fields in de Sitter $\R^{1,4}$ and anti-de Sitter
$\R^{3,2}$ spaces. Importance of discrete transformations is well-known, many textbooks
on quantum theory began with description of the discrete symmetries, and
famous L\"{u}ders-Pauli $CPT$ Theorem is a keystone in general structure of quantum field
theory (see, for example, the excellent book \cite{SW64}). Besides, a fundamental notion of antimatter immediately relates with the charge conjugation $C$. However, usual practice of definition of the discrete symmetries
from the analysis of relativistic wave equations does not give a full and
consistent theory of the discrete transformations. In the standard approach,
except a well studied case of the spin $j=1/2$ (Dirac equation), a situation
with the discrete symmetries in quantum field theory remains unclear for the fields of higher spin
$j>1/2$. It is obvious that the main reason of this is an absence of a fully
adequate formalism for description of higher-spin fields (all widely
accepted higher-spin formalisms such as Rarita--Schwinger approach \cite{RS41},
Bargmann-Wigner \cite{BW48} and Gel'fand-Yaglom \cite{GY48} multispinor
theories, and also Joos-Weinberg $2(2j+1)$-component formalism
\cite{Joo62,Wein} have many intrinsic contradictions and difficulties).
Moreover, Lee and Wick \cite{LW66} claimed that ``the situation is clearly
an unsatisfactory one from a fundamental point of view".
The first attempt of going out from this situation was initiated by
Gel'fand, Minlos and Shapiro in 1958 \cite{GMS}. In the
Gel'fand-Minlos-Shapiro approach the discrete symmetries are represented
by outer involutory automorphisms of the Lorentz group (there are also other
realizations of the discrete symmetries via the outer automorphisms, see
\cite{Mic64,Kuo71,Sil92}).
At present, the Gel'fand-Minlos-Shapiro ideas have been found further development in the
works of Buchbinder, Gitman and Shelepin \cite{BGS00,GS01}, where the
discrete symmetries are represented by both outer and inner automorphisms of
the Poincar\'{e} group. An algebraic method for description of discrete symmetries was
proposed by author in the works \cite{Var01,Var03},
where the discrete symmetries are represented by
automorphisms of the Clifford algebras.

In this paper $CPT$ groups are considered in the spaces $\R^{4,1}$
and $\R^{1,4}$ with mutually opposite signatures $(+,+,+,+,-)$ and
$(-,-,-,-,+)$. The difference of signatures induces difference of
Clifford algebras associated with these spaces. So, for the algebra
$\cl_{4,1}$, associated with the space $\R^{4,1}$, there is an
isomorphism $\cl_{4,1}\simeq\cl_4$, where $\cl_4$ is a Dirac algebra,
and for the algebra $\cl_{1,4}$, associated with the space
$\R^{1,4}$, we have a decomposition
$\cl_{1,4}\simeq\cl_{4,0}\oplus\cl_{4,0}$, where
$\cl_{4,0}\simeq\cl^+_{1,4}$ is an even subalgebra of $\cl_{1,4}$.
$CPT$ groups of the algebra $\cl_{4,1}\simeq\cl_4$ are studied in the
section 3 with respect to all real subalgebras of $\cl_4$. In turn,
for the algebra $\cl_{1,4}$ there exists a homomorphic mapping
$\epsilon:\;\cl_{1,4}\rightarrow{}^\epsilon\cl_{4,0}$, where
${}^\epsilon\cl_{4,0}\simeq\cl_{4,0}/\Ker\epsilon$ is a quotient
algebra, $\Ker\epsilon$ is the kernel of the homomorphism $\epsilon$.
It is shown that at the mapping $\epsilon$ the $CPT$ group of the
initial algebra is reduced to one from its subgroups. Discrete
symmetries on the quotient representations of the algebra
$\cl_{1,4}$, associated with the space $\R^{1,4}$, are studied in
the section 4. The analogous situation takes place for the algebra
$\cl_{3,2}$ associated with the anti-de Sitter space $\R^{3,2}$. In
this case we have a decomposition
$\cl_{3,2}\simeq\cl_{2,2}\oplus\cl_{2,2}$, where
$\cl_{2,2}\simeq\cl^+_{3,2}$ is an even subalgebra of $\cl_{3,2}$ (this decomposition is a particular
case of the well-known isomorphism $\cl_{p,q}\simeq\cl_{q,p-1}\oplus\cl_{q,p-1}$, see, for example, \cite{BT88}).
Discrete symmetries of the quotient algebra ${}^\epsilon\cl_{2,2}$,
obtained at the homomorphic mapping
$\epsilon:\;\cl_{3,2}\rightarrow{}^\epsilon\cl_{2,2}$, are
considered in the section 5.
\section{$CPT$ group}
As is known, within the Clifford algebras there are infinitely many (continuous) automorphisms.
Discrete symmetries $P$ and $T$ transform (reflect) space and time
(two the most fundamental notions in physics), but in the Minkowski
4-dimensional spacetime continuum space and time are not
separate and independent. For that reason a transformation of one (space or
time) induces a transformation of another. Therefore, discrete symmetries
should be expressed by such transformations of the continuum, which transform
all its structure totally with a full preservation of discrete nature\footnote{It is interesting to note that in the well-known Penrose twistor program \cite{Pen} a spinor structure is understood as the underlying (more fundamental) structure with respect to Minkowski spacetime. In other words, spacetime continuum is not fundamental substance in the twistor approach, this is a fully derivative (in spirit of Leibnitz philosophy) entity generated by the underlying spinor structure. In this context spacetime discrete symmetries $P$ and $T$ should be considered as projections (shadows) of the fundamental automorphisms belonging to the background spinor structure.}. In 1949, Schouten \cite{Sch49} introduced such
(discrete) automorphisms. In 1955, a first systematic description of these automorphisms was given by Rashevskii
\cite{Ras55}. He showed that within the Clifford algebra $\cl_{p,q}$ over the real field $\F=\R$ there exist \emph{four fundamental automorphisms}:\\[0.2cm]
1) {\bf Identity}: An automorphism $\cA\rightarrow\cA$ and
$\e_{i}\rightarrow\e_{i}$.\\
This automorphism is an identical automorphism of the algebra $\cl_{p,q}$.
$\cA$ is an arbitrary element of $\cl_{p,q}$.\\[0.2cm]
2) {\bf Involution}: An automorphism $\cA\rightarrow\cA^\star$ and
$\e_{i}\rightarrow-\e_{i}$.\\
In more details, for an arbitrary element $\cA\in\cl_{p,q}$ there exists a
decomposition
$
\cA=\cA^{\p}+\cA^{\p\p},
$
where $\cA^{\p}$ is an element consisting of homogeneous odd elements, and
$\cA^{\p\p}$ is an element consisting of homogeneous even elements,
respectively. Then the automorphism
$\cA\rightarrow\cA^{\star}$ is such that the element
$\cA^{\p\p}$ is not changed, and the element $\cA^{\p}$ changes sign:
$
\cA^{\star}=-\cA^{\p}+\cA^{\p\p}.
$
If $\cA$ is a homogeneous element, then
\begin{equation}\label{auto16}
\cA^{\star}=(-1)^{k}\cA,
\end{equation}
where $k$ is a degree of the element. It is easy to see that the
automorphism $\cA\rightarrow\cA^{\star}$ may be expressed via the volume
element $\omega=\e_{12\ldots p+q}$:
\begin{equation}\label{auto17}
\cA^{\star}=\omega\cA\omega^{-1},
\end{equation}
where
$\omega^{-1}=(-1)^{\frac{(p+q)(p+q-1)}{2}}\omega$. When $k$ is odd, the basis
elements
$\e_{i_{1}i_{2}\ldots i_{k}}$ the sign changes, and when $k$ is even, the sign
is not changed.\\[0.2cm]
3) {\bf Reversion}: An antiautomorphism $\cA\rightarrow\widetilde{\cA}$ and
$\e_i\rightarrow\e_i$.\\
The antiautomorphism $\cA\rightarrow\widetilde{\cA}$ is a reversion of the
element $\cA$, that is the substitution of each basis element
$\e_{i_{1}i_{2}\ldots i_{k}}\in\cA$ by the element
$\e_{i_{k}i_{k-1}\ldots i_{1}}$:
\[
\e_{i_{k}i_{k-1}\ldots i_{1}}=(-1)^{\frac{k(k-1)}{2}}
\e_{i_{1}i_{2}\ldots i_{k}}.
\]
Therefore, for any $\cA\in\cl_{p,q}$ we have
\begin{equation}\label{auto19}
\widetilde{\cA}=(-1)^{\frac{k(k-1)}{2}}\cA.
\end{equation}
4) {\bf Conjugation}: An antiautomorphism $\cA\rightarrow\widetilde{\cA^\star}$
and $\e_i\rightarrow-\e_i$.\\
This antiautomorphism is a composition of the antiautomorphism
$\cA\rightarrow\widetilde{\cA}$ with the automorphism
$\cA\rightarrow\cA^{\star}$. In the case of a homogeneous element from
the formulae (\ref{auto16}) and (\ref{auto19}), it follows
\begin{equation}\label{20}
\widetilde{\cA^{\star}}=(-1)^{\frac{k(k+1)}{2}}\cA.
\end{equation}
A finite group structure of the
automorphism set
$\{\Id,\,\star,\,\widetilde{\phantom{cc}},\,\widetilde{\star}\}$ was
studied in \cite{Var01} with respect to discrete symmetries which
compound $PT$ group (so-called \emph{reflection group})\footnote{Some applications of the fundamental automorphisms to discrete symmetries of quantum field theory were considered by Rashevskii in \cite{Ras55} (see also his paper \cite{Ras58})}.

Other important discrete symmetry is the charge conjugation $C$. In contrast
with the transformations $P$, $T$, $PT$, the operation $C$ is not
spacetime discrete symmetry. As is known, the Clifford algebra $\cl_n$ over the complex field $\F=\C$ is associated with a complex vector
space $\C^n$. Let $n=p+q$, then an extraction operation of the real subspace
$\R^{p,q}$ in $\C^n$  forms the foundation of definition of the discrete
transformation known in physics as
{\it a charge conjugation} $C$. Indeed, let
$\{\e_1,\ldots,\e_n\}$ be an orthobasis in the space $\C^n$, $\e^2_i=1$.
Let us remain the first $p$ vectors of this basis unchanged, and other $q$
vectors multiply by the factor $i$. Then the basis
\begin{equation}\label{6.23}
\left\{\e_1,\ldots,\e_p,i\e_{p+1},\ldots,i\e_{p+q}\right\}
\end{equation}
allows one to extract the subspace $\R^{p,q}$ in $\C^n$. Namely,
for the vectors $\R^{p,q}$ we take the vectors of
$\C^n$ which decompose on the basis
(\ref{6.23}) with real coefficients. In such a way we obtain a real vector
space $\R^{p,q}$ endowed (in general case) with a non-degenerate
quadratic form
\[
Q(x)=x^2_1+x^2_2+\ldots+x^2_p-x^2_{p+1}-x^2_{p+2}-\ldots-x^2_{p+q},
\]
where $x_1,\ldots,x_{p+q}$ are coordinates of the vector $\bx$
in the basis (\ref{6.23}).
It is easy to see that the extraction of
$\R^{p,q}$ in $\C^n$ induces an extraction of
{\it a real subalgebra}
$\cl_{p,q}$ in $\cl_n$. Therefore, any element
$\cA\in\cl_n$ can be unambiguously represented in the form
\[
\cA=\cA_1+i\cA_2,
\]
where $\cA_1,\,\cA_2\in\cl_{p,q}$. The one-to-one mapping
\begin{equation}\label{6.24}
\cA\longrightarrow\overline{\cA}=\cA_1-i\cA_2
\end{equation}
transforms the algebra $\cl_n$ into itself with preservation of addition
and multiplication operations for the elements $\cA$; the operation of
multiplication of the element $\cA$ by the number transforms to an operation
of multiplication by the complex conjugate number.
Any mapping of $\cl_n$ satisfying these conditions is called
{\it a pseudoautomorphism}.
Thus, the extraction of the subspace
$\R^{p,q}$ in the space $\C^n$ induces in the algebra $\cl_n$
a pseudoautomorphism $\cA\rightarrow\overline{\cA}$ \cite{Ras55,Ras58}. Compositions of $\cA\rightarrow\overline{\cA}$
with the fundamental automorphisms allow one to extend the set
$\{\Id,\,\star,\,\widetilde{\phantom{cc}},\,\widetilde{\star}\}$ by
the pseudoautomorphisms $\cA\rightarrow\overline{\cA}$,
$\cA\rightarrow\overline{\cA^\star}$,
$\cA\rightarrow\overline{\widetilde{\cA}}$,
$\cA\rightarrow\overline{\widetilde{\cA^\star}}$ \cite{Var03}. A
finite group structure of \emph{an automorphism set}
$\{\Id,\,\star,\,\widetilde{\phantom{cc}},\,\widetilde{\star},\,
\overline{\phantom{cc}},\,\overline{\star},\,
\overline{\widetilde{\phantom{cc}}},\,\overline{\widetilde{\star}}\}$
was studied in \cite{Var03} with respect to $CPT$ symmetries.

Further, in accordance with Wedderburn theorem any finite-dimensional associative simple algebra $\mathfrak{A}$ over the field $\F$ is isomorphic to a full matrix algebra $\Mat_n(\K)$, where $n$ is
natural number defined unambiguously, and $\K$ a division ring defined
with an accuracy of isomorphism.
According to Wedderburn theorem
the antiautomorphism $\cA\rightarrow\widetilde{\cA}$ corresponds to
an antiautomorphism of the full matrix algebra $\Mat_{2^m}(\K)$:
$\sA\rightarrow\sA^{t}$, in virtue of the well-known relation
$(\sA\sB)^{t}=
\sB^{t}\sA^{t}$, where $t$ is a symbol of transposition, $n=2m$. On the other hand,
in the matrix representation of the elements $\cA\in\cl_{p,q}$, for the
antiautomorphism
$\cA\rightarrow\widetilde{\cA}$ we have $\sA\rightarrow\widetilde{\sA}$.
A composition of the two antiautomorphisms, $\sA^{t}\rightarrow\sA\rightarrow
\widetilde{\sA}$, gives an automorphism  $\sA^{t}\rightarrow\widetilde{\sA}$,
which is an internal automorphism of the algebra $\Mat_{2^m}(\K)$:
\begin{equation}\label{com}
\widetilde{\sA}=\sE\sA^{t}\sE^{-1},
\end{equation}
where $\sE$ is a matrix, by means of which the antiautomorphism $\cA\rightarrow
\widetilde{\cA}$ is expressed in the matrix representation of the
algebra $\cl_{p,q}$.
Under action of the antiautomorphism $\cA\rightarrow\widetilde{\cA}$
the units of $\cl_{p,q}$ remain unaltered, $\e_i\rightarrow\e_i$; therefore
in the matrix representation, we must demand $\cE_i\rightarrow\cE_i$,
where $\cE_i=\gamma(\e_i)$ also.
Therefore, for the definition of the matrix $\sE$
in accordance with (\ref{com}) we have
\begin{equation}\label{com1}
\cE_i\longrightarrow\cE_i=\sE\cE^{t}\sE^{-1}.
\end{equation}
A spinor representation of the automorphism $\cA\rightarrow\cA^\star$ follows directly from (\ref{auto17}): $\sA^\star=\sW\sA\sW^{-1}$, where $\sW=\cE_1\cE_2\cdots\cE_{p+q}$. Further,
a spinor representation of the pseudoautomorphism
$\cA\rightarrow\overline{\cA}$ of the algebra $\cl_n$ when $n\equiv 0\s\pmod{2}$
is defined as follows.
In the spinor representation the every element $\cA\in\cl_n$ should be
represented by some matrix $\sA$, and the pseudoautomorphism (\ref{6.24})
takes a form of the pseudoautomorphism of the full
matrix algebra
$\Mat_{2^{n/2}}$:
\[
\sA\longrightarrow\overline{\sA}.
\]\begin{sloppypar}\noindent
On the other hand, a transformation replacing the matrix $\sA$ by the
complex conjugate matrix, $\sA\rightarrow\dot{\sA}$, is also some
pseudoautomorphism of the algebra $\Mat_{2^{n/2}}$. The composition of the two
pseudoautomorpisms $\dot{\sA}\rightarrow\sA$ and
$\sA\rightarrow\overline{\sA}$, $\dot{\sA}\rightarrow\sA\rightarrow
\overline{\sA}$, is an internal automorphism
$\dot{\sA}\rightarrow\overline{\sA}$ of the full matrix algebra $\Mat_{2^{n/2}}$:
\end{sloppypar}
\begin{equation}\label{6.25}
\overline{\sA}=\Pi\dot{\sA}\Pi^{-1},
\end{equation}
where $\Pi$ is a matrix of the pseudoautomorphism
$\cA\rightarrow\overline{\cA}$ in the spinor representation.
The sufficient condition for definition of the pseudoautomorphism
$\cA\rightarrow\overline{\cA}$ is a choice of the matrix
$\Pi$ in such a way that the transformation
$\sA\rightarrow\Pi\dot{\sA}\Pi^{-1}$ transfers into itself the matrices
$\cE_1,\ldots,\cE_p,i\cE_{p+1},\ldots,i\cE_{p+q}$
(the matrices of the spinbasis of $\cl_{p,q}$), that is,
\begin{equation}\label{6.26}
\cE_i\longrightarrow\cE_i=\Pi\dot{\cE}_i\Pi^{-1}\quad
(i=1,\ldots,p+q).
\end{equation}
The more detailed spinor representation of the pseudoautomorphism $\cA\rightarrow\overline{\cA}$ with respect to a division ring structure of the real subalgebras is given by the following theorem.
\begin{theorem}[{\rm\cite{Var03}}]\label{tpseudo}
Let $\cl_n$ be a complex Clifford algebra for $n\equiv 0\s\pmod{2}$
and let $\cl_{p,q}\subset\C_n$ be its subalgebra with a real division ring
$\K\simeq\R$ when $p-q\equiv 0,2\s\pmod{8}$ and quaternionic division ring
$\K\simeq\BH$ when $p-q\equiv 4,6\s\pmod{8}$, $n=p+q$. Then in dependence
on the division ring structure of the real subalgebra $\cl_{p,q}$ the matrix
$\Pi$ of the pseudoautomorphism $\cA\rightarrow\overline{\cA}$
has the following form:\\[0.2cm]
1) $\K\simeq\R$, $p-q\equiv 0,2\s\pmod{8}$.\\[0.1cm]
The matrix $\Pi$ for any spinor representation over the ring $\K\simeq\R$
is proportional to the unit matrix.\\[0.2cm]
2) $\K\simeq\BH$, $p-q\equiv 4,6\s\pmod{8}$.\\[0.1cm]
$\Pi=\cE_{\alpha_1\alpha_2\cdots\alpha_a}$ when
$a\equiv 0\s\pmod{2}$ and
$\Pi=\cE_{\beta_1\beta_2\cdots\beta_b}$ when $b\equiv 1\s\pmod{2}$,
where $a$ complex matrices $\cE_{\alpha_t}$
and $b$ real matrices $\cE_{\beta_s}$ form a basis of the spinor
representation of the algebra $\cl_{p,q}$ over the ring $\K\simeq\BH$,
$a+b=p+q,\,0<t\leq a,\,0<s\leq b$. At this point,
\begin{eqnarray}
\Pi\dot{\Pi}&=&\phantom{-}\sI\quad\text{if $a,b\equiv 0,1\s\pmod{4}$},
\nonumber\\
\Pi\dot{\Pi}&=&-\sI\quad\text{if $a,b\equiv 2,3\s\pmod{4}$},\nonumber
\end{eqnarray}
where $\sI$ is the unit matrix.
\end{theorem}
Spinor representations of the all other automorphisms from the set $\{\Id,\,\star,\,\widetilde{\phantom{cc}},\,\widetilde{\star},\,
\overline{\phantom{cc}},\,\overline{\star},\,
\overline{\widetilde{\phantom{cc}}},\,\overline{\widetilde{\star}}\}$ are defined in a similar manner.
We list these transformations and their spinor representations:
\begin{eqnarray}
\cA\longrightarrow\cA^\star,&&\quad\sA^\star=\sW\sA\sW^{-1},\nonumber\\
\cA\longrightarrow\widetilde{\cA},&&\quad\widetilde{\cA}=\sE\sA^{t}\sE^{-1},
\nonumber\\
\cA\longrightarrow\widetilde{\cA^\star},&&\quad\widetilde{\sA^\star}=
\sC\sA^{t}\sC^{-1},\quad\sC=\sE\sW,\nonumber\\
\cA\longrightarrow\overline{\cA},&&\quad\overline{\sA}=\Pi\sA^\ast\Pi^{-1},
\nonumber\\
\cA\longrightarrow\overline{\cA^\star},&&\quad\overline{\sA^\star}=
\sK\sA^\ast\sK^{-1},\quad\sK=\Pi\sW,\nonumber\\
\cA\longrightarrow\overline{\widetilde{\cA}},&&\quad
\overline{\widetilde{\sA}}=\sS\left(\sA^{t}\right)^\ast\sS^{-1},\quad
\sS=\Pi\sE,\nonumber\\
\cA\longrightarrow\overline{\widetilde{\cA^\star}},&&\quad
\overline{\widetilde{\sA^\star}}=\sF\left(\sA^\ast\right)^{t}\sF^{-1},\quad
\sF=\Pi\sC,\nonumber
\end{eqnarray}

It is easy to verify that an automorphism
set $\{\Id,\,\star,\,\widetilde{\phantom{cc}},\,\widetilde{\star},\,
\overline{\phantom{cc}},\,\overline{\star},\,
\overline{\widetilde{\phantom{cc}}},\,\overline{\widetilde{\star}}\}$
of $\cl_n$ forms a finite group of order 8.
\begin{sloppypar}
Further, let $\cl_n$ be a Clifford algebra over the field $\F=\C$ and
let $\Ext(\cl_n)=
\{\Id,\,\star,\,\widetilde{\phantom{cc}},\,\widetilde{\star},\,
\overline{\phantom{cc}},\,\overline{\star},\,
\overline{\widetilde{\phantom{cc}}},\,\overline{\widetilde{\star}}\}$
be \emph{an automorphism group} of the algebra $\cl_n$. Then
there is an isomorphism between $\Ext(\cl_n)$ and a $CPT$ group of
the discrete transformations,
$\Ext(\cl_n)\simeq\{1,\,P,\,T,\,PT,\,C,\,CP,\,CT,\,CPT\}\simeq
\dZ_2\times\dZ_2\times\dZ_2$. In this case, space inversion $P$,
time reversal $T$, full reflection $PT$, charge conjugation $C$,
transformations $CP$, $CT$ and the full $CPT$--transformation
correspond to the automorphism $\cA\rightarrow\cA^\star$,
antiautomorphisms $\cA\rightarrow\widetilde{\cA}$,
$\cA\rightarrow\widetilde{\cA^\star}$, pseudoautomorphisms
$\cA\rightarrow\overline{\cA}$,
$\cA\rightarrow\overline{\cA^\star}$, pseudoantiautomorphisms
$\cA\rightarrow\overline{\widetilde{\cA}}$ and
$\cA\rightarrow\overline{\widetilde{\cA^\star}}$, respectively
\cite{Var03}.\end{sloppypar}
\begin{sloppypar}
The group $\{1,\,P,\,T,\,PT,\,C,\,CP,\,CT,\,CPT\}$ at the conditions
$P^2=T^2=(PT)^2=C^2=(CP)^2=(CT)^2=(CPT)^2=1$ and commutativity of
all the elements forms an Abelian group of order 8, which is
isomorphic to a cyclic group $\dZ_2\times\dZ_2\times\dZ_2$. In
turn, the automorphism group
$\{\Id,\,\star,\,\widetilde{\phantom{cc}},\,\widetilde{\star},\,
\overline{\phantom{cc}},\,\overline{\star},\,
\overline{\widetilde{\phantom{cc}}},\,\overline{\widetilde{\star}}\}$
in virtue of commutativity $\widetilde{\left(\cA^\star\right)}=
\left(\widetilde{\cA}\right)^\star$,
$\overline{\left(\cA^\star\right)}=\left(\overline{\cA}\right)^\star$,
$\overline{\left(\widetilde{\cA}\right)}=
\widetilde{\left(\overline{\cA}\right)}$,
$\overline{\left(\widetilde{\cA^\star}\right)}=
\widetilde{\left(\overline{\cA}\right)^\star}$ and an involution
property
$\star\star=\widetilde{\phantom{cc}}\widetilde{\phantom{cc}}=
\overline{\phantom{cc}}\;\overline{\phantom{cc}}=\Id$ is also
isomorphic to $\dZ_2\times\dZ_2\times\dZ_2$:\end{sloppypar}
\[
\{1,\,P,\,T,\,PT,\,C,\,CP,\,CT,\,CPT\}\simeq
\{\Id,\,\star,\,\widetilde{\phantom{cc}},\,\widetilde{\star},\,
\overline{\phantom{cc}},\,\overline{\star},\,
\overline{\widetilde{\phantom{cc}}},\,\overline{\widetilde{\star}}\}\simeq
\dZ_2\times\dZ_2\times\dZ_2.
\]
In 2003, the $CPT$ group was
introduced \cite{Var03} in the context of an extension of
automorphism groups of Clifford algebras. The relationship between
$CPT$ groups and extraspecial groups and universal coverings of
orthogonal groups was established in \cite{Var03,Var04}. In 2004,
Socolovsky considered the $CPT$ group of the spinor field with
respect to phase factors \cite{Soc04}. $CPT$ groups of spinor fields
in the de Sitter spaces of different signatures were studied in the
works \cite{Var05c,Var05}. $CPT$ groups for higher spin fields have
been defined in \cite{Var11} on the spinspaces associated with
representations of the spinor group $\spin_+(1,3)$ (a universal
covering of the proper orthocronous Lorentz group).
\subsection{Salingaros groups}
As it has been shown previously, $CPT$ groups have explicit realizations via the finite groups.
As is known \cite{Sal81a}--\cite{Sha94}, a structure of the Clifford
algebras admits a very elegant description in terms of finite groups.
In accordance
with a multiplication rule
\begin{equation}\label{e1}
\e^2_i=\sigma(p-i)\e_0,\quad\e_i\e_j=-\e_j\e_i,
\end{equation}
\begin{equation}\label{e2}
\sigma(n)=\left\{\begin{array}{rl}
-1 & \mbox{if $n\leq 0$},\\
+1 & \mbox{if $n>0$},
\end{array}\right.
\end{equation}
basis elements  of the Clifford algebra $\cl_{p,q}$ (the algebra
over the field of real numbers, $\F=\R$) form a finite group of
order $2^{n+1}$,
\begin{equation}\label{FG}
G(p,q)=\left\{\pm
1,\,\pm\e_i,\,\pm\e_i\e_j,\,\pm\e_i\e_j\e_k,\,\ldots,\,
\pm\e_1\e_2\cdots\e_n\right\}\quad(i<j<k<\ldots).
\end{equation}
Salingaros showed \cite{Sal81a,Sal82} that there exist five distinct types
of finite groups (\ref{FG}) that arise from Clifford algebras.
In \cite{Sal81a,Sal82} they were called `vee groups' and were labelled as
\begin{equation}\label{FG2}
N_{\text{odd}},\;N_{\text{even}},\;\Omega_{\text{odd}},\;\Omega_{\text{even}},\;
S_k.
\end{equation}
The odd $N$-groups correspond to real spinors, for example, $N_1$ is related
to real 2-spinors, and $N_3$ is the group of the real Majorana matrices.
The even $N$-groups define the quaternionic groups.
The $S$-groups
are the `spinor groups'
($S_k=N_{2k}\times\C\simeq N_{2k-1}\times\C$):
$S_1$ is the group of the complex Pauli matrices, and $S_2$ is the group of the
Dirac matrices. Furthermore, the $\Omega$-groups are double copies of the
$N$-groups and can be written as a direct product of the $N$-groups with
the group of two elements $\dZ_2$:
\begin{equation}\label{FG3}
\Omega_k=N_k\times\dZ_2.
\end{equation}
The five
distinct types of Salingaros groups correspond to the five division rings
of the real Clifford algebras \cite{Var04}:
\begin{eqnarray}
N_{\text{odd}}&\leftrightarrow&\cl_{p,q},\;p-q\equiv 0,2\pmod{8},\;\K\simeq\R;
\nonumber\\
N_{\text{even}}&\leftrightarrow&\cl_{p,q},\;p-q\equiv 4,6\pmod{8},\;
\K\simeq\BH;\nonumber\\
\Omega_{\text{odd}}&\leftrightarrow&\cl_{p,q},\;p-q\equiv 1\pmod{8},\;
\K\simeq\R\oplus\R;\nonumber\\
\Omega_{\text{even}}&\leftrightarrow&\cl_{p,q},\;p-q\equiv 5\pmod{8},\;
\K\simeq\BH\oplus\BH;\nonumber\\
S_k&\leftrightarrow&\cl_{p,q},\;p-q\equiv 3,7\pmod{8},\;\K\simeq\C.\nonumber
\end{eqnarray}
Let $\bZ(p,q)\subset\cl_{p,q}$ be a center of the finite group (\ref{FG}).
In such a way, we have three distinct realizations of the center
$\bZ(p,q)$:
\begin{eqnarray}
\bZ(p,q)&=&\{1,-1\}\simeq\dZ_2\;\;\text{if}\;p-q\equiv 0,2,4,6\pmod{8};
\nonumber\\
\bZ(p,q)&=&\{1,-1,\omega,-\omega\}\simeq\dZ_2\times\dZ_2\;\;\text{if}\;
p-q\equiv 1,5\pmod{8};\nonumber\\
\bZ(p,q)&=&\{1,-1,\omega,-\omega\}\simeq\dZ_4\;\;\text{if}\;
p-q\equiv 3,7\pmod{8}.\nonumber
\end{eqnarray}
The Abelian groups $\bZ(p,q)$ are the subgroups of the Salingaros groups
(\ref{FG2}). Namely, $N$--groups have the center $\dZ_2$,
$\Omega$--groups have the center $\dZ_2\times\dZ_2$, and $S$-group has
the center $\dZ_4$.

The following Theorem presents a key result in the group structure of
$\cl_{p,q}$.
\begin{theorem}[{\rm Salingaros \cite{Sal81a}}]
The factor group $G(p,q)/\bZ(p,q)$ is the Abelian group
$\left(\dZ_2\right)^{\times 2k}=\dZ_2\times\dZ_2\times\cdots\times\dZ_2$
($2k$ times):
\[
\frac{G(p,q)}{\bZ(p,q)}:\;\;
\frac{N_{2k-1}}{\dZ_2}\simeq\frac{N_{2k}}{\dZ_2}\simeq
\frac{\Omega_{2k-1}}{\dZ_2\times\dZ_2}\simeq
\frac{\Omega_{2k}}{\dZ_2\times\dZ_2}\simeq
\frac{S_k}{\dZ_4}\simeq\left(\dZ_2\right)^{\times 2k}.
\]
\end{theorem}
This Theorem allows one to identify the Salingaros groups with extraspecial
groups \cite{Sal81a,Sal84,Bra85}.
As known, a finite group $G$ is called
an extraspecial 2-group if $G$ is of order $2^n$ and $G/\bZ(G)$ is the
Abelian group $\dZ_2\times\cdots\times\dZ_2$ ($n-1$ times). Further, if
$G$ is the extraspecial 2-group of order $2^{2k+1}$, then
\begin{eqnarray}
G&\simeq&D_4\circ\cdots\circ D_4\quad (k\;\,\text{times}),\;\;\text{or}
\nonumber\\
G&\simeq&Q_4\circ D_4\circ\cdots\circ D_4\quad (k-1\;\,\text{times}),
\nonumber
\end{eqnarray}
where $\circ$ means a \emph{central product}
of two groups: that is, the direct
product with centers identified. For example, the direct product
$Q_4\times D_4$ has the resulting group of order $8\times 8=64$, and its
center is the direct product
of the two individual centers and is equal to
$\dZ_2\times\dZ_2$. In contrast, the central product $Q_4\circ D_4$
amalgamates the $\dZ_2$ center of $Q_4$ with the $\dZ_2$ of $D_4$ to give
the center of $Q_4\circ D_4$ as $\dZ_2$. Therefore, the central product
$Q_4\circ D_4$ is of order $32$. In the case where one center is a subgroup
of the other center, they both amalgamate into the larger center.

In terms of the extraspecial groups all the Salingaros groups take the form:
\begin{eqnarray}
N_{2k-1}&\simeq&\left(N_1\right)^{\circ k},\nonumber\\
N_{2k}&\simeq&N_2\circ\left(N_1\right)^{\circ(k-1)},\nonumber\\
\Omega_{2k-1}&\simeq&N_{2k-1}\circ\left(\dZ_2\times\dZ_2\right),\nonumber\\
\Omega_{2k}&\simeq&N_{2k}\circ\left(\dZ_2\times\dZ_2\right),\nonumber\\
S_k&\simeq&N_{2k-1}\circ\dZ_4\simeq N_{2k}\circ\dZ_4.\nonumber
\end{eqnarray}

As is known, an important property of each finite group is its order
structure. The order of a particular element $\alpha$ in the group
is the smallest integer $p$ for which $\alpha^p=1$. For example, the
Tab.\,1 lists the number of distinct elements in each group which
have order 2, 4, or 8 (the identity 1 is the only element of order
1).
\begin{figure}[ht]
\begin{center}{\renewcommand{\arraystretch}{1.3}
\begin{tabular}{|l|c|ccc|}\hline
   &  & \multicolumn{3}{l}{Order structure}\vline\\
   & Type & 2 & 4 & 8 \\ \hline
1. $\dZ_2\times\dZ_2\times\dZ_2$ & Abelian & 7 & & \\
2. $\dZ_4\times\dZ_2$ & & 3 & 4 & \\
3. $\dZ_8$ & & 1 & 2 & 4\\ \hline
4. $D_4$ & Non--Abelian & 5 & 2 & \\
5. $Q_4$ & & 1 & 6 & \\
6. $\overset{\ast}{\dZ}_4\times\dZ_2$ & & 3 & 4 &\\ \hline
\end{tabular}
}
\end{center}
\hspace{0.3cm}
\begin{center}{\small
{\bf Tab.\,1:} Finite groups of order 8.}
\end{center}
\end{figure}

Following to Salingaros an \emph{order structure of the finite group} is defined by the expression
\[
(o_1,\,o_2,\,o_4,\,o_8),
\]
where $o_1$ is the number of elements of order 1, $o_2$ is the number of elements of order 2 and so on. Salingaros drops the first entry as it always 1. Therefore, we will use below the order structure with three entries, $(o_2,\,o_4,\,o_8)$. Of course, $\dZ_8$ does not occur as a $G(p,q)$ (Salingaros group),
since every element of $G(p,q)$ has order 1, 2, or 4. The groups
$\dZ_4\times\dZ_2$ and $\overset{\ast}{\dZ}_4\times\dZ_2$ in Tab.\,1 have the
same order structure (3,4,0), but the sequences of `+' and `-' are different in their signatures $(a,b,c,d,e,f,g)$. The group $\overset{\ast}{\dZ}_4\times\dZ_2$ is an non-Abelian analogue of $\dZ_4\times\dZ_2$.
\subsection{D\c{a}browski groups}
In 1958, Shirokov pointed out \cite{Shi58,Shi60} that a universal covering of the
inhomogeneous Lorentz group has eight inequivalent realizations. Later on,
in the eighties this idea was applied to a general orthogonal group
$O(p,q)$ by D\c{a}browski \cite{Dab88}.
As is known, the orthogonal
group $O(p,q)$ of the real space $\R^{p,q}$ is represented by the semidirect
product of a connected component $O_0(p,q)$ and a discrete subgroup
$\{1,P,T,PT\}$. In general,
there are eight double coverings of the orthogonal group
$O(p,q)$ \cite{Dab88}:
\[
\rho^{a,b,c}:\;\;\pin^{a,b,c}(p,q)\longrightarrow O(p,q),
\]\begin{sloppypar}\noindent
where $a,b,c\in\{+,-\}$. As is known, the group $O(p,q)$ consists of four
connected components: identity connected component $O_0(p,q)$, and three
components corresponding to space inversion $P$, time reversal
$T$, and the combination of these two $PT$, that is, $O(p,q)=(O_0(p,q))\cup
P(Q_0(p,q))\cup T(O_0(p,q))\cup PT(O_0(p,q))$. Further, since the
four-element group (reflection group) $\{1,\,P,\,T,\,PT\}$ is isomorphic to
the finite group $\dZ_2\times\dZ_2$, then
$O(p,q)$ may be represented by
a semidirect product $O(p,q)\simeq O_0(p,q)
\odot(\dZ_2\otimes\dZ_2)$. The signs of $a,b,c$ correspond to the signs of the
squares of the elements in $\pin^{a,b,c}(p,q)$ which cover space inversion
$P$, time reversal $T$ and a combination of these two
$PT$ ($a=-P^2,\,b=T^2,\,c=-(PT)^2$ in D\c{a}browski's notation \cite{Dab88} and
$a=P^2,\,b=T^2,\,c=(PT)^2$ in Chamblin's notation \cite{Ch94} which we will
use below).
An explicit form of the group $\pin^{a,b,c}(p,q)$ is given by the following
semidirect product:\end{sloppypar}
\begin{equation}\label{Pinabc}
\pin^{a,b,c}(p,q)\simeq\frac{(\spin_+(p,q)\odot C^{a,b,c})}{\dZ_2},
\end{equation}
where $C^{a,b,c}$ are the four double coverings of
$\dZ_2\times\dZ_2$.
All the eight universal coverings of the orthogonal group
$O(p,q)$ are given in the Tab.\,2.
\begin{figure}[ht]
\begin{center}
{\renewcommand{\arraystretch}{1.3}
\begin{tabular}{|c|l|l|}\hline
$a$ $b$ $c$ & $C^{a,b,c}$ & Remark \\ \hline
$+$ $+$ $+$ & $\dZ_2\times\dZ_2\times\dZ_2$ & $PT=TP$\\
$+$ $-$ $-$ & $\dZ_2\times\dZ_4$ & $PT=TP$\\
$-$ $+$ $-$ & $\dZ_2\times\dZ_4$ & $PT=TP$\\
$-$ $-$ $+$ & $\dZ_2\times\dZ_4$ & $PT=TP$\\ \hline
$-$ $-$ $-$ & $Q_4$ & $PT=-TP$\\
$-$ $+$ $+$ & $D_4$ & $PT=-TP$\\
$+$ $-$ $+$ & $D_4$ & $PT=-TP$\\
$+$ $+$ $-$ & $D_4$ & $PT=-TP$\\ \hline
\end{tabular}
}
\end{center}
\hspace{0.3cm}
\begin{center}{\small
{\bf Tab.\,2:} $PT$-structures.}
\end{center}
\end{figure}
At this point, the group
\[
\Aut(\cl_{p,q})\simeq\frac{C^{a,b,c}}{\dZ_2}
\]
is the reflection group.

In turn, it has been shown \cite{Var03} that there exist 64 universal
coverings of the orthogonal group $O(p,q)$:
\[
\rho^{a,b,c,d,e,f,g}:\;\pin^{a,b,c,d,e,f,g}\longrightarrow O(p,q),
\]
where
\begin{equation}\label{RCL}
\pin^{a,b,c,d,e,f,g}(p,q)\simeq\frac{(\spin_+(p,q)\odot
C^{a,b,c,d,e,f,g})}{\dZ_2}.
\end{equation}
Here $C^{a,b,c,d,e,f,g}$ are five double coverings of the group
$\dZ_2\times\dZ_2\times\dZ_2$, and $a,b,c,d,e,f,g\in\{+,-\}$, $a=P^2$, $b=T^2$, $c=(PT)^2$, $d=C^2$, $e=(CP)^2$, $f=(CT)^2$, $g=(CPT)^2$. All the possible double coverings
$C^{a,b,c,d,e,f,g}$ are given in the Tab.\,3.
\begin{figure}[ht]
\begin{center}{\renewcommand{\arraystretch}{1.3}
\begin{tabular}{|l|l|l|}\hline
$a\;b\;c\;d\;e\;f\;g$ & $C^{a,b,c,d,e,f,g}$ & Type \\ \hline
$+\;+\;+\;+\;+\;+\;+$ & $\dZ_2\times\dZ_2\times\dZ_2\times\dZ_2$ & Abelian \\
three `$+$' and four `$-$' & $\dZ_4\times\dZ_2\times\dZ_2$ & \\ \hline
one `$+$' and six `$-$' & $Q_4\times\dZ_2$ & Non-Abelian \\
five `$+$' and two `$-$' & $D_4\times\dZ_2$ &   \\
three `$+$' and four `$-$' & $\overset{\ast}{\dZ}_4\times\dZ_2\times\dZ_2$
& \\ \hline
\end{tabular}
}
\end{center}
\hspace{0.3cm}
\begin{center}{\small
{\bf Tab.\,3:} $CPT$-structures.}
\end{center}
\end{figure}
Thus,
\[
C^{a,b,c,d,e,f,g}=\{\pm 1,\,\pm P,\,\pm T,\,\pm PT,\,\pm C,\,\pm
CP,\, \pm CT,\,\pm CPT\}
\]
is {\it a full $CPT$ group}.
$C^{a,b,c,d,e,f,g}$ is a finite group of order 16 (a complete
classification of these groups is given in \cite{Var03}). At this
point, the group
\[
\Ext(\cl_{p,q})=\frac{C^{a,b,c,d,e,f,g}}{\dZ_2}
\]
is called {\it a generating group}. It is easy to see that in case of the algebra $\cl_{p,q}$ (or subalgebra
$\cl_{p,q}\subset\cl_n$) with the real division ring $\K\simeq\R$,
$p-q\equiv 0,2\pmod{8}$, $CPT$-structures, defined by the group
(\ref{RCL}), are reduced to the eight
Shirokov-D\c{a}browski $PT$-structures.

\section{$CPT$ groups in the space $\R^{4,1}$}
In 1935, Dirac \cite{Dir35} introduced relativistic wave equations
in a five-dimensional pseudoeuclidean space  $\R^{4,1}$ (de Sitter
space),
\begin{equation}\label{DirSit}
(i\gamma_0\partial_0+i\gamma_k\partial_k-m)\psi=0
\end{equation}
or
\[
(i\gamma_\mu\partial_\mu+m)\psi=0,
\]
where five $4\times 4$ Dirac matrices $\gamma_\mu$ satisfy the
relations
\[
\gamma_\mu\gamma_\nu+\gamma_\nu\gamma_\mu=2g_{\mu\nu},\quad
\mu=0,1,2,3,4.
\]
The algebra $\cl_{4,1}$, associated with the space $\R^{4,1}$, has the type $p-q\equiv 3\pmod{8}$.
Hence it follows the isomorphism $\cl_{4,1}\simeq\cl_4$, where $\cl_4$
is a Clifford algebra over the complex field $\F=\C$ (so-called
\emph{Dirac algebra}).

A finite group $G(4,1)$ corresponding to $\cl_{4,1}\simeq\cl_4$ is a
particular case of (\ref{FG}). So, in accordance with (\ref{FG}) the
{\it Dirac group} $G(4,1)$ is defined by the following set:
\begin{multline}
G(4,1)=\{\pm1,\,\pm\e_1,\,\ldots,\,\pm\e_5,\,
\pm\e_1\e_2,\,\ldots,\,\pm\e_4\e_5,\\
\pm\e_1\e_2\e_3,\,\ldots,\,\pm\e_3\e_4\e_5,\,\pm\e_1\e_2\e_3\e_4,\,
\ldots,\,\pm\e_2\e_3\e_4\e_5,\,\pm\e_1\e_2\e_3\e_4\e_5\}. \nonumber
\end{multline}
It is a finite group of order 64 with the order structure $(31,32,0)$.
The Dirac group is an extraspecial two-group. For this group the
following isomorphism holds:
\[
G(4,1)=S_2\simeq N_4\circ\dZ_4\simeq Q_4\circ D_4\circ\dZ_4.
\]
The center of $G(4,1)$ is isomorphic to
the group $\dZ_4$ (finite group of order 4). $G(4,1)$ is an
non-Abelian group (as all Salingaros groups, except the first three
groups $\dZ_2$, $\Omega_0=\dZ_2\times\dZ_2$ and $S_0=\dZ_4$).

In dependence on the division ring structure the algebra $\cl_4$ has
five real subalgebras which correspond to real subspaces
$\R^{p,q}\subset\C^4$ ($p+q=4$) with distinct signatures
$(+,-,-,-)$, $(-,+,+,+)$, $(-,-,-,-)$, $(+,+,+,+)$ and $(-,-,+,+)$.
Three subalgebras with the quaternionic ring $\K\simeq\BH$: the
spacetime algebra $\cl_{1,3}$, $\cl_{4,0}$ and $\cl_{0,4}$. Two
subalgebras with the real ring $\K\simeq\R$: the {\it Majorana}
$\cl_{3,1}$ and $\cl_{2,2}$ algebras.
In accordance with (\ref{FG}) the each real subalgebra
$\cl_{p,q}\subset\C_4$ induces a finite group $G(p,q)$.
Let us consider in detail the structure of these finite groups.
Owing to (\ref{FG}), a
{\it spacetime group} is defined by the followng set:
\begin{multline}
G(1,3)=\{\pm 1,\,\pm\gamma_0,\,\pm\gamma_1,\,\pm\gamma_2,\,
\pm\gamma_3,\,\pm\gamma_0\gamma_1,\,
\pm\gamma_0\gamma_2,\\
\pm\gamma_0\gamma_3,\,
\pm\gamma_1\gamma_2,\,\pm\gamma_1\gamma_3,\,
\pm\gamma_2\gamma_3,\,\pm\gamma_0\gamma_1\gamma_2,\,
\pm\gamma_0\gamma_1\gamma_3,\,\\
\pm\gamma_0\gamma_2\gamma_3,\,\pm\gamma_1\gamma_2\gamma_3,\,
\pm\gamma_0\gamma_1\gamma_2\gamma_3\}.\label{DG}
\end{multline}
It is a finite group of order 32
with the order structure
$(11,20,0)$. Moreover,
$G(1,3)$ is the extraspecial two-group.
In Salingaros notation the following isomorphism holds:
\[
G(1,3)=N_4\simeq Q_4\circ D_4,
\]
A center of the group $G(1,3)$ is isomorphic to a cyclic group
$\dZ_2$. $G(1,3)$ is the non-Abelian group which contains many subgroups both
Abelian and non-Abelian.
For example, the group of fundamental automorphisms of the algebra
$\cl_{1,3}$ is an Abelian subgroup of
$G(1,3)$, $\Aut(\cl_{1,3})=\{\Id,\star,
\widetilde{\phantom{cc}},\widetilde{\star}\}\simeq\{1,P,T,PT\}
\simeq\dZ_2\times\dZ_2
\subset G(1,3)$. In turn,
the automorphism group of the algebra
$\cl_{1,3}$ is a non-Abelian subgroup of
$G(1,3)$, $\Ext(\cl_{1,3})\simeq\subset G(1,3)$. Finite groups $G(4,0)$ and $G(0,4)$, corresponding to the subalgebras
$\cl_{4,0}$ and $\cl_{0,4}$, are isomorphic to each
other, since these groups possess the order structure
$(11,20,0)$. This group isomorphism is a direct consequence of the algebra
isomorphism
$\cl_{4,0}\simeq\cl_{0,4}$. Therefore,
\[
G(4,0)\simeq G(0,4)=N_4\simeq Q_4\circ D_4.
\]
The Majorana group $G(3,1)$ with the order structure $(19,12,0)$ is a central
product of the two groups $D_4$:
\[
G(3,1)=N_3\simeq N_1\circ N_1\simeq D_4\circ D_4.
\]
$G(3,1)$ is the non-Abelian group; a center of the group is isomorphic to
$\dZ_2$.
The same isomorphism takes place for the group
\[
G(2,2)=N_3\simeq D_4\circ D_4.
\]
Thus, the real subalgebras of the Dirac algebra $\cl_4$ form five finite
groups of order 32. Three subalgebras with the ring
$\K\simeq\BH$ form finite groups which isomorphic to the central product
$Q_4\circ D_4$, and two subalgebras with the ring $\K\simeq\R$ form finite
groups defined by the product $D_4\circ D_4$.

If we consider a spinor representation (a left regular
representation in a spinspace $\dS$), then the units $\e_i$ of the
algebra $\cl_4$ are replaced by $\gamma$-matrices via the rule
$\gamma_i=\gamma(\e_i)$, where $\gamma$ is a mapping of the form
$\cl_{p,q}\overset{\gamma}{\longrightarrow}\End_{\K}(\dS)$,
$\cl_{p,q}\subset\cl_4$. In the papers \cite{Var05c,Var05} matrices
$\gamma_i=\gamma(\e_i)$ were defined via {\it a Brauer-Weyl
representation} \cite{BW35}. Following to a more rigorous algebraic
framework \cite{Lou91}, we see that a definition procedure of the
spinor representation over the ring $\K$ is hardly fixed and depends
on the structure of primitive idempotents $f$ of the algebra
$\cl_{p,q}$. As is known, for the Clifford algebra $\cl_{p,q}$ over
the field $\F=\R$ there are isomorphisms
$\cl_{p,q}\simeq\End_{\K}(I_{p,q})\simeq\Mat_{2^m}(\K)$, where
$m=(p+q)/2$, $I_{p,q}=\cl_{p,q}f$ is a minimal left ideal of
$\cl_{p,q}$, and $\K=f\cl_{p,q}f$ is a division ring of $\cl_{p,q}$.
A primitive idempotent of the algebra $\cl_{p,q}$ has the form
\[
f=\frac{1}{2}(1\pm\e_{\alpha_1})\frac{1}{2}(1\pm\e_{\alpha_2})\cdots\frac{1}{2}
(1\pm\e_{\alpha_k}),
\]
where $\e_{\alpha_1},\e_{\alpha_2},\ldots,\e_{\alpha_k}$ are
commuting elements with square 1 of the canonical basis of
$\cl_{p,q}$ generating a group of order $2^k$, that is,
$(\e_{\alpha_1},\e_{\alpha_2},\ldots,\e_{\alpha_k})\simeq\left(\dZ_2\right)^{\times
k}$, where $\left(\dZ_2\right)^{\times
k}=\dZ_2\times\dZ_2\times\cdots\times\dZ_2$ ($k$ times) is an
Abelian group. The values of $k$ are defined by the formula
$k=q-r_{q-p}$, where $r_i$ are the Radon-Hurwitz numbers
\cite{Rad22,Hur23}, values of which form a cycle of the period 8:
$r_{i+8}=r_i+4$. The values of all $r_i$ are
\begin{center}
\begin{tabular}{lcccccccc}
$i$  & 0 & 1 & 2 & 3 & 4 & 5 & 6 & 7\\ \hline $r_i$& 0 & 1 & 2 & 2 &
3 & 3 & 3 & 3
\end{tabular}.
\end{center}\begin{sloppypar}\noindent
In terms of finite groups we have here \emph{an idempotent group}
$T_{p,q}(f)\simeq\left(\dZ_2\right)^{\times (k+1)}$ of the order
$2^{k+1}=2^{1+q-r_{q-p}}$. In turn, a quotient group
$G(p,q)/T_{p,q}(f)\simeq G_{p,q}(f)$ is a finite group of order
$2^{1+p+r_{q-p}}$ \cite{AF01,AF02,AF03}. It is obvious that
$G_{p,q}(f)$ is \emph{a normal subgroup} of $G(p,q)$.\end{sloppypar}

First of all, in accordance with Theorem 9 in \cite{Var03} Clifford
algebras over the field $\F=\C$ correspond to \textbf{\emph{charged
particles}} such as electron, proton and so on. $CPT$ groups for the algebra $\cl_{4,1}\simeq\cl_4$ are considered in appendix.

In general case all the elements of $C^{a,b,c,d,e,f,g}$ (resp.
$\sExt(\cl_{1,3})$) depend on the phase factors. Let us suppose
\begin{equation}\label{GenCPT}
P=\eta_p\sW,\quad T=\eta_t\sE,\quad C=\eta_c\Pi,
\end{equation}
where $\eta_p,\,\eta_t,\,\eta_c\in\C^\ast=\C-\{0\}$. Taking into
account (\ref{GenCPT}), we obtain
\begin{multline}
\sExt(\cl_{1,3})\simeq\{1,\,P,\,T,\,PT,\,C,\,CP,\,CT,\,CPT\}\simeq\\
\simeq\{\boldsymbol{1}_4,\,\eta_p\sW,\,\eta_t\sE,\,\eta_p\eta_t\sE\sW,\,\eta_c\Pi,\,\eta_c\eta_p\Pi\sW,\,
\eta_c\eta_t\Pi\sE,\,\eta_c\eta_p\eta_t\Pi\sE\sW\}\simeq\\
\simeq\{\boldsymbol{1}_4,\,\eta_p\sW,\,\eta_t\sE,\,\eta_p\eta_t\sC,\,\eta_c\Pi,\,\eta_c\eta_p\sK,\,
\eta_c\eta_t\sS,\,\eta_c\eta_p\eta_t\sF\}.\nonumber
\end{multline}
The multiplication table of this general group is given in Tab.\,4.
\begin{figure}[ht]
{\scriptsize
\begin{center}{\renewcommand{\arraystretch}{1.6}
\begin{tabular}{|c||c|c|c|c|c|c|c|c|}\hline
  & $\boldsymbol{1}_4$ & $\eta_p\sW$ & $\eta_t\sE$ & $\eta_{pt}\sC$ & $\eta_c\Pi$ &
$\eta_{cp}\sK$ & $\eta_{ct}\sS$ & $\eta_{cpt}\sF$\\
\hline\hline $\boldsymbol{1}_4$ & $\boldsymbol{1}_4$ & $\eta_p\sW$ &
$\eta_t\sE$ &
$\eta_{pt}\sC$ & $\eta_c\Pi$ & $\eta_{cp}\sK$ & $\eta_{ct}\sS$ & $\eta_{cpt}\sF$\\
\hline $\eta_p\sW$ & $\eta_p\sW$ & $\eta^2_p\sW^2$ &
$\eta_{pt}\sW\sE$ &
$\eta^2_p\eta_t\sW\sC$ & $\eta_{pc}\sW\Pi$ & $\eta_c\eta^2_p\sW\sK$ & $\eta_{cpt}\sW\sS$ & $\eta_{ct}\eta^2_p\sW\sF$\\
\hline $\eta_t\sE$ & $\eta_t\sE$ & $\eta_{pt}\sE\sW$ &
$\eta^2_t\sE^2$ & $\eta_p\eta^2_t\sE\sC$
& $\eta_{ct}\sE\Pi$ & $\eta_{cpt}\sE\sK$ & $\eta_c\eta^2_t\sE\sS$ & $\eta_{cp}\eta^2_t\sE\sF$\\
\hline $\eta_{pt}\sC$ & $\eta_{pt}\sC$ & $\eta^2_p\eta_t\sC\sW$ &
$\eta_p\eta^2_t\sC\sE$ &
 $\eta^2_{pt}\sC^2$ & $\eta_{cpt}\sC\Pi$ & $\eta_{ct}\eta^2_p\sC\sK$ & $\eta_{cp}\eta^2_t\sC\sS$ &
$\eta_c\eta^2_{pt}\sC\sF$\\ \hline $\eta_c\Pi$ & $\eta_c\Pi$ &
$\eta_{cp}\Pi\sW$ & $\eta_{ct}\Pi\sE$ & $\eta_{cpt}\Pi\sC$ &
$\eta^2_c\Pi^2$ & $\eta^2_c\eta_p\Pi\sK$ & $\eta^2_c\eta_t\Pi\sS$ & $\eta^2_c\eta_{pt}\Pi\sF$\\
\hline $\eta_{cp}\sK$ & $\eta_{cp}\sK$ & $\eta_c\eta^2_p\sK\sW$ &
$\eta_{cpt}\sK\sE$ & $\eta_{ct}\eta^2_p\sK\sC$
& $\eta^2_c\eta_p\sK\Pi$ & $\eta^2_{cp}\sK^2$ & $\eta^2_c\eta_{pt}\sK\sS$ & $\eta^2_{cp}\eta_t\sK\sF$\\
\hline $\eta_{ct}\sS$ & $\eta_{ct}\sS$ & $\eta_{cpt}\sS\sW$ &
$\eta_c\eta^2_t\sS\sE$ & $\eta_{cp}\eta^2_t\sS\sC$ &
$\eta^2_c\eta_t\sS\Pi$ & $\eta^2_c\eta_{pt}\sS\sK$ &
$\eta^2_{ct}\sS^2$ & $\eta^2_{ct}\eta_p\sS\sF$\\
\hline $\eta_{cpt}\sF$ & $\eta_{cpt}\sF$ & $\eta_{ct}\eta^2_p\sF\sW$
& $\eta_{cp}\eta^2_t\sF\sE$ & $\eta_c\eta^2_{pt}\sF\sC$ &
$\eta^2_c\eta_{pt}\sF\Pi$ & $\eta^2_{cp}\eta_t\sF\sK$ &
$\eta^2_{ct}\eta_p\sF\sS$ & $\eta^2_{cpt}\sF^2$\\
\hline
\end{tabular}
}
\end{center}
} \hspace{0.3cm}
\begin{center}\begin{minipage}{22pc}
{\small \textbf{Tab.\,4:} The multiplication table of general
generating group $\sExt(\cl_{1,3})$.}
\end{minipage}
\end{center}
\end{figure}
The Tab.\,4 presents \emph{a general generating matrix} for any
possible $CPT$ groups of the fields of any spin. However, for the
spinor field $\boldsymbol{\xi}^\alpha$ (spin-1/2 field) there are
some restrictions. As is known, many textbooks on quantum field
theory (see, for example, \cite{Sch,Capr}) state that a fermion and
its associated antifermion have \emph{opposite} relative intrinsic
parity, that is, for the field of type $(1/2,0)\oplus(0,1/2)$ charge
conjugation $C$ and space inversion $P$ anticommute, $CP=-PC$,
$\{C,P\}=0$. At first glance, for higher spin fields, defined
via tensor products
$\boldsymbol{\xi}^{\alpha_1\alpha_2\ldots\alpha_k}=\sum\boldsymbol{\xi}^{\alpha_1}\otimes
\boldsymbol{\xi}^{\alpha_2}\otimes\cdots\otimes\boldsymbol{\xi}^{\alpha_k}$,
these restrictions should be changed. Hence it follows that
\textit{bosonic} and \textit{fermionic} $CPT$ groups should be
different. Indeed, a boson and its associated antiboson carry
\emph{same} relative intrinsic parity, that is, for the field of
type $(1,0)\oplus(0,1)$ the operations $C$ and $P$ commute, $CP=PC$,
$[C,P]=0$, see \cite{Wein}. However, Wigner \cite{Wig64} showed that
there are theories where a fermion and its associated antifermion
have the \emph{same} relative intrinsic parity, $[C,P]=0$; and that a
boson and its associated antiboson carry \emph{opposite} relative
intrinsic parity, $\{C,P\}=0$. Later on such theories were called as
Bargmann-Wightman-Wigner-type quantum field theories
\cite{Ahl93,Ahl95}, see also \cite{Dvo09,Dvo12,RR05} and references
therein. At present day this ambiguous situation is known as
`$[C,P]_\pm=0$ dilemma' \cite{Dvo12}.

In context of the present theory of discrete symmetries
`$[C,P]_\pm=0$ dilemma' has a simple algebraic solution. It is
obvious that a choice $[C,P]_\pm=0$ depends on the spinor
representations of $P=\eta_p\sW$ and $C=\eta_c\Pi$. In general, the
matrix $\Pi$ has two different forms:
$\Pi=\gamma_{\alpha_1}\gamma_{\alpha_2}\cdots\gamma_{\alpha_a}$ when
$a\equiv 0\pmod{2}$ and
$\Pi=\gamma_{\beta_1}\gamma_{\beta_2}\cdots\gamma_{\beta_b}$ when
$b\equiv 1\pmod{2}$, where the complex matrices $\gamma_{\alpha_t}$
and the real matrices $\gamma_{\beta_s}$ form a basis of the spinor
representation of the algebra $\cl_{p,q}$ over the ring
$\K\simeq\BH$, $a+b=p+q$, $0<t\leq a$, $0<s\leq b$ \cite{Var03}. It
is easy to verify that
$\Pi=\gamma_{\alpha_1}\gamma_{\alpha_2}\cdots\gamma_{\alpha_a}$
always commutes with $\sW$, therefore, we have $[C,P]=0$ in this
case. In turn,
$\Pi=\gamma_{\beta_1}\gamma_{\beta_2}\cdots\gamma_{\beta_b}$ always
anticommutes with $\sW$, therefore, $\{C,P\}=0$. Among the groups
$\sExt(\cl_{1,3})$, considered in this subsection, we see that
$\sExt^+_1(\cl_{1,3})$ with $\Pi=\gamma_{23}$,
$\sExt^+_2(\cl_{1,3})$ and $\sExt^+_3(\cl_{1,3})$ with
$\Pi=\gamma_{34}$ and $\sExt^+_4(\cl_{1,3})$ with $\Pi=\gamma_{24}$
lead to $[C,P]=0$. In contrast, $\sExt^+_5(\cl_{1,3})$ with
$\Pi=\gamma_1$ leads to $\{C,P\}=0$.

\section{$CPT$ groups in the space $\R^{1,4}$}
In the work \cite{MRT05} discrete symmetries for the spinor field in
the de Sitter space with the signature $(+,-,-,-,-)$, that is, in
the space $\R^{1,4}$, have been derived via the analysis of a de
Sitter-Dirac wave equation. Discrete symmetries in the de Sitter
space with the signature $(+,+,+,+,-)$, that is, in the space
$\R^{4,1}$, have been considered in the previous section within an
algebraic framework based on the automorphism set of Clifford
algebras. In this section we study group structure of discrete
transformations in the space $\R^{1,4}$.

In this context the de Sitter spacetime is understood as a
hyperboloid embedded in the space $\R^{1,4}$:
\begin{equation}\label{dS}
X_H=\left\{x\in\R^{1,4}:\;x^2=\eta_{\alpha\beta}x^\alpha
x^\beta=-H^{-2}\right\},\quad \alpha,\beta=0,1,2,3,4,
\end{equation}
where $\eta_{\alpha\beta}=\text{diag}(1,-1,-1,-1,-1)$.

The spinor wave equation in the Sitter spacetime (\ref{dS}) has been
derived in \cite{BGMT01,MRT05}. This equation has the form
\begin{equation}\label{dSDir}
(-i\not\! x\gamma\cdot\overline{\partial}+2i+\nu)\psi(x)=0,
\end{equation}
where $\not\! x=\eta^{\alpha\beta}\gamma_\alpha x_\beta$ and
$\overline{\partial}_\alpha=\partial_\alpha+H^2x_\alpha
x\cdot\partial$. In this case $4\times 4$ matrices $\gamma_\alpha$
are spinor representations of $\cl_{1,4}$ and satisfy the relations
\[
\gamma_\alpha\gamma_\beta+\gamma_\beta\gamma_\alpha=2\eta_{\alpha\beta}
\boldsymbol{1}_4.
\]

The finite group $G(1,4)$, associated with the algebra $\cl_{1,4}$,
is a particular case of (\ref{FG}). It is a finite group of order 64
with the order structure (23,40,0). In Salingaros notation we have the
following isomorphism:
\[
G(1,4)=\Omega_4\simeq N_4\circ(\dZ_2\times\dZ_2)\simeq Q_4\circ
D_4\circ(\dZ_2\times\dZ_2).
\]

Let us define $CPT$ group for the spinor field in the space
$\R^{1,4}$. First of all, the transformation $C$ (the
pseudoautomorphism $\cA\rightarrow\overline{\cA}$) for the algebras
$\cl_{p,q}$ over the field $\F=\R$ and the ring $\K\simeq\BH$ (the
types $p-q\equiv 4,6\pmod{8}$) corresponds to
\textbf{\emph{particle-antiparticle interchange}} $C^\prime$ (see
\cite{Var03,Var04}). As is known, neutral particles are described
within real representations of the Lorentz group. There are two
classes of neutral particles: 1) particles which have antiparticles
such as neutrons, neutrinos and so on; 2) particles which coincide
with their antiparticles (for example, photons). The first class is
described by the the algebras $\cl_{p,q}$ over the field $\F=\R$
with the rings $\K\simeq\BH$ and $\K\simeq\BH\oplus\BH$ (the types
$p-q\equiv 4,6\pmod{8}$ and $p-q\equiv 5\pmod{8}$), and the second
class (\textbf{\emph{truly neutral particles}}) is described by the algebras
$\cl_{p,q}$ over the field $\F=\R$ with the rings $\K\simeq\R$ and
$\K\simeq\R\oplus\R$ (the types $p-q\equiv 0,2\pmod{8}$ and
$p-q\equiv 1\pmod{8}$) (for more details see
\cite{Var03,Var04,Var12}).

Since $\cl_{1,4}$ is a semisimple algebra over the field $\F=\R$,
then for $\cl_{1,4}$ we have the following decomposition:
$\cl_{1,4}\simeq\cl^+_{1,4}\oplus\cl^+_{1,4}$, where $\cl^+_{1,4}$
is an even subalgebra of $\cl_{1,4}$. Moreover, there is an
isomorphism $\cl^+_{1,4}\simeq\cl_{4,0}$, therefore,
$\cl_{1,4}\simeq\cl_{4,0}\oplus\cl_{4,0}$. This decomposition can be
represented by the following scheme:
\[
\unitlength=0.5mm
\begin{picture}(70,50)
\put(35,40){\vector(2,-3){15}} \put(35,40){\vector(-2,-3){15}}
\put(28.25,42){$\cl_{1,4}$} \put(16,28){$\lambda_{+}$}
\put(49.5,28){$\lambda_{-}$} \put(9.5,9.20){$\cl_{4,0}$}
\put(47.75,9){$\cl_{4,0}$} \put(32.5,10){$\oplus$}
\end{picture}
\]
Here central idempotents
\[
\lambda_+=\frac{1+\e_1\e_2\e_3\e_4\e_5}{2},\quad\lambda_-=\frac{1-\e_1\e_2\e_3\e_4\e_5}{2}
\]
satisfy the relations $(\lambda_+)^2=\lambda_+$,
$(\lambda_-)^2=\lambda_-$, $\lambda_+\lambda_-=0$. In accordance
with \cite{CF96} the idempotents $\lambda_+$ and $\lambda_-$ can be
identified with \emph{helicity projection operators} which
distinguish left and right handed spinors.

The decomposition $\cl_{1,4}\simeq\cl_{4,0}\oplus\cl_{4,0}$ induces
a left-regular spinor representation
$\cl_{1,4}\overset{\gamma}{\longrightarrow}\End_{\BH\oplus\hat{\BH}}
(\dS_2\oplus\hat{\dS}_2)$, where $\dS_2(\BH)\simeq
I_{4,0}=\cl_{4,0}f$ is a minimal left ideal of the subalgebra
$\cl_{4,0}$, $f$ is a primitive idempotent of $\cl_{4,0}$. This
spinor representation is realized within the matrix algebra
${}^2\Mat_2(\BH)=\Mat_2(\BH)\oplus\Mat_2(\BH)$.

However, there is a homomorphic mapping
\[
\epsilon:\;\cl_{1,4}\longrightarrow{}^\epsilon\cl_{4,0},
\]
where
\[
{}^\epsilon\cl_{4,0}\simeq\cl_{4,0}/\Ker\epsilon
\]
is a quotient algebra,
$\Ker\epsilon=\left\{\cA^1-\omega\cA^1\right\}$ is a kernel of the
homomorphism $\epsilon$ (a bilateral ideal of the algebra
$\cl_{1,4}$, see \cite{Var04}), $\cA^1\in\cl_{4,0}$,
$\omega=\e_{12345}$ is a volume element of $\cl_{1,4}$. Under action
of the homomorphism $\epsilon$ we cannot to transfer all the
automorphisms of the initial algebra $\cl_{1,4}$ onto the quotient
algebra ${}^\epsilon\cl_{4,0}$. So, in accordance with Theorem 14 in
\cite{Var04} at the mapping
$\epsilon:\;\cl_{1,4}\rightarrow{}^\epsilon\cl_{4,0}$ we can
transfer onto ${}^\epsilon\cl_{4,0}$ only the antiautomorphism
$\cA\rightarrow\widetilde{\cA}$, pseudoautomorphism
$\cA\rightarrow\overline{\cA}$ (particle-antiparticle interchange
$C^\prime$) and pseudoantiautomorphism
$\cA\rightarrow\overline{\widetilde{\cA}}$. All other automorphisms
of $\cl_{1,4}$ do not transferred from $\cl_{1,4}$ to
${}^\epsilon\cl_{4,0}$. Therefore, we have
$\sAut({}^\epsilon\cl_{4,0})=\{1,T,C^\prime,C^\prime
T\}\sim\{\sI,\sE,\Pi,\sS\}$. It is easy to see that
$\{1,T,C^\prime,C^\prime T\}\sim\{\sI,\sE,\Pi,\sS\}$ is the subgroup
of the group $\sExt(\cl_{4,0})$ considered in the subsection 3.2.
For the idempotents $f^\pm_1$, $f^\pm_2$, $f^\pm_3$,
$f^\pm_4\in\cl_{4,0}$, which generate the group (\ref{Group^+_1}),
we obtain
\[
\sAut({}^\epsilon\cl_{4,0})\simeq\{1,T,C^\prime,C^\prime
T\}\simeq\{\boldsymbol{1}_4,\gamma_{34},\eta_c\gamma_{34},\eta_c\boldsymbol{1}_4\}.
\]
The multiplication table of this group is shown in the Tab.\,5.
\begin{figure}[ht]
{\footnotesize
\[
{\renewcommand{\arraystretch}{1.4}
\begin{tabular}{|c||c|c|c|c|}\hline
   & $1$ & $T$ & $C^\prime$ & $C^\prime T$ \\ \hline\hline
$1$& $1$ & $T$ & $C^\prime$ & $C^\prime T$ \\ \hline $T$& $T$ & $-1$ & $-C^\prime T$ & $C^\prime$ \\
\hline $C^\prime$ & $C^\prime$ & $-C^\prime T$ & $-(C^\prime)^2$ & $(C^\prime)^2T$ \\
\hline $C^\prime T$ & $C^\prime T$ & $C^\prime$ & $(C^\prime)^2T$ & $(C^\prime T)^2$ \\
\hline
\end{tabular}\sim
\begin{tabular}{|c||c|c|c|c|}\hline
   & $\boldsymbol{1}_4$ & $\gamma_{34}$ & $\eta_c\gamma_{34}$ &
$\eta_c\boldsymbol{1}_{4}$ \\ \hline\hline $\boldsymbol{1}_4$ &
$\boldsymbol{1}_4$ & $\gamma_{34}$ & $\eta_c\gamma_{34}$ &
$\eta_c\boldsymbol{1}_{4}$
\\ \hline $\gamma_{34}$ & $\gamma_{34}$ &
$-\boldsymbol{1}_4$ &
$-\eta_c\boldsymbol{1}_{4}$ & $\eta_c\gamma_{34}$ \\
\hline $\eta_c\gamma_{34}$ & $\eta_c\gamma_{34}$ &
$-\eta_c\boldsymbol{1}_{4}$ & $-\eta^2_c\boldsymbol{1}_{4}$ &
$\eta^2_c\gamma_{34}$
\\ \hline $\eta_c\boldsymbol{1}_{4}$ &
$\eta_c\boldsymbol{1}_{4}$ & $\eta_c\gamma_{34}$ &
$\eta^2_c\gamma_{34}$ & $\eta^2_c\boldsymbol{1}_{4}$
\\ \hline
\end{tabular}
}
\]
} \hspace{0.3cm}
\begin{center}\begin{minipage}{20pc}
{\small \textbf{Tab.\,5:} The multiplication table of
$\sAut({}^\epsilon\cl_{4,0})\simeq\dZ_4$ in dependence of the sign
of $\eta^2_c$. $\dZ_4$ with the signature $(-,-,+)$ at $\eta^2_c=1$
and $\dZ_4$ with $(-,+,-)$ at $\eta^2_c=-1$.}
\end{minipage}
\end{center}
\end{figure}
In both cases $\eta^2_c=1$ and $\eta^2_c=-1$ we have here an Abelian
group $\sAut_-({}^\epsilon\cl_{4,0})\simeq\dZ_4$ and, therefore,
$C^\prime PT$ groups (more precisely, $C^\prime T$ groups) are
isomorphic to $C^{-,-,+}\simeq C^{-,+,-}\simeq\dZ_4\times\dZ_2$.

Further, for the idempotent
$f^+_5=\frac{1}{2}(1+\e_{1234})\in\cl_{4,0}$, which generates the
group (\ref{Group^+_5}), we have
\[
\sAut^+_5({}^\epsilon\cl_{4,0})\simeq\{1,T,C^\prime,C^\prime
T\}\simeq\{\boldsymbol{1}_4,\gamma_{1},\eta_c\gamma_{1},\eta_c\boldsymbol{1}_4\}.
\]
The multiplication table of this group is given in Tab.\,6.
\begin{figure}[ht]
\[
{\renewcommand{\arraystretch}{1.4}
\begin{tabular}{|c||c|c|c|c|}\hline
   & $1$ & $T$ & $C^\prime$ & $C^\prime T$ \\ \hline\hline
$1$& $1$ & $T$ & $C^\prime$ & $C^\prime T$ \\ \hline $T$& $T$ & $1$ & $C^\prime T$ & $C^\prime$ \\
\hline $C^\prime$ & $C^\prime$ & $C^\prime T$ & $(C^\prime)^2$ & $(C^\prime)^2T$ \\
\hline $C^\prime T$ & $C^\prime T$ & $C^\prime$ & $(C^\prime)^2T$ & $(C^\prime T)^2$ \\
\hline
\end{tabular}\,\sim\,
\begin{tabular}{|c||c|c|c|c|}\hline
   & $\boldsymbol{1}_4$ & $\gamma_{1}$ & $\eta_c\gamma_{1}$ &
$\eta_c\boldsymbol{1}_{4}$ \\ \hline\hline $\boldsymbol{1}_4$ &
$\boldsymbol{1}_4$ & $\gamma_{1}$ & $\eta_c\gamma_{1}$ &
$\eta_c\boldsymbol{1}_{4}$
\\ \hline $\gamma_{1}$ & $\gamma_{1}$ &
$\boldsymbol{1}_4$ &
$\eta_c\boldsymbol{1}_{4}$ & $\eta_c\gamma_{1}$ \\
\hline $\eta_c\gamma_{1}$ & $\eta_c\gamma_{1}$ &
$\eta_c\boldsymbol{1}_{4}$ & $\eta^2_c\boldsymbol{1}_{4}$ &
$\eta^2_c\gamma_{1}$
\\ \hline $\eta_c\boldsymbol{1}_{4}$ &
$\eta_c\boldsymbol{1}_{4}$ & $\eta_c\gamma_{1}$ &
$\eta^2_c\gamma_{1}$ & $\eta^2_c\boldsymbol{1}_{4}$
\\ \hline
\end{tabular}
}
\]
\hspace{0.3cm}
\begin{center}\begin{minipage}{20pc}
{\small \textbf{Tab.\,6:} The multiplication table of
$\sAut^+_5({}^\epsilon\cl_{4,0})$ in dependence on the sign of
$\eta^2_c$. $\sAut^+_5({}^\epsilon\cl_{4,0})\simeq\dZ_2\times\dZ_2$
with the signature $(+,+,+)$ at $\eta^2_c=1$ and
$\sAut^+_5({}^\epsilon\cl_{4,0})\simeq\dZ_4$ with $(+,-,-)$ at
$\eta^2_c=-1$.}
\end{minipage}
\end{center}
\end{figure}
In this case when $\eta^2_c=1$ we have the group $\sAut^+_5({}^\epsilon\cl_{4,0})\simeq\dZ_2\times\dZ_2$
and the group $\sAut^+_5({}^\epsilon\cl_{4,0})\simeq\dZ_4$
when $\eta^2_c=-1$. In the first case we come to
$C^{+,+,+}\simeq\dZ_2\times\dZ_2\times\dZ_2$ and to
$C^{+,-,-}\simeq\dZ_4\times\dZ_2$ in the second case.

\section{$CPT$ groups in anti-de Sitter space $\R^{3,2}$}
As it has been shown in the section 4, a five-dimensional
pseudoeuclidean space $\R^{1,4}$ (so called {\it de Sitter space})
is associated with the algebra $\cl_{1,4}$. In turn, the {\it
anti-de Sitter space} $\R^{3,2}$, associated with the algebra
$\cl_{3,2}$, leads to the following extraspecial group of order 64:
\[
G(3,2)=\Omega_3\simeq N_3\circ D_2\simeq D_4\circ D_4\circ D_2.
\]

Let us study discrete symmetries arising from the algebra
$\cl_{3,2}$ associated with the space $\R^{3,2}$. First of all,
$\cl_{3,2}$ is a semisimple algebra over the field $\F=\R$ with a
double division ring $\K\simeq\R\oplus\R$, the type $p-q\equiv
1\pmod{8}$. In accordance with Theorem 9 in \cite{Var03} for the
algebras $\cl_{p,q}$ over the real field $\F=\R$ with real division
rings $\K\simeq\R$ and $\K\simeq\R\oplus\R$ (types $p-q\equiv
0,1,2\pmod{8}$) the pseudoautomorphism
$\cA\rightarrow\overline{\cA}$ (charge conjugation $C$) is reduced
to the identical transformation $\Id$, therefore, such algebras
correspond to \textbf{\emph{truly neutral particles}} (for example, photons,
$K^0$-mesons and so on).

The algebra $\cl_{3,2}$ admits a decomposition
$\cl_{3,2}\simeq\cl^+_{3,2}\oplus\cl^+_{3,2}$, where
$\cl^+_{3,2}\simeq\cl_{2,2}$ is an even subalgebra of $\cl_{3,2}$.
In turn, this decomposition can be represented by the scheme
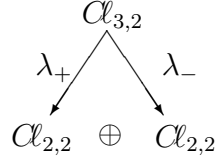
\[
\unitlength=0.5mm
\begin{picture}(70,50)
\put(35,40){\vector(2,-3){15}} \put(35,40){\vector(-2,-3){15}}
\put(28.25,42){$\cl_{3,2}$} \put(16,28){$\lambda_{+}$}
\put(49.5,28){$\lambda_{-}$} \put(9.5,9.20){$\cl_{2,2}$}
\put(47.75,9){$\cl_{2,2}$} \put(32.5,10){$\oplus$}
\end{picture}
\]
Here central idempotents
\[
\lambda_+=\frac{1+\e_1\e_2\e_3\e_4\e_5}{2},\quad\lambda_-=\frac{1-\e_1\e_2\e_3\e_4\e_5}{2}
\]
generate two mutually annihilating simple ideals.

Further, the decomposition $\cl_{3,2}\simeq\cl_{2,2}\oplus\cl_{2,2}$
induces a left-regular spinor representation
$\cl_{3,2}\overset{\gamma}{\longrightarrow}\End_{\R\oplus\hat{\R}}
(\dS_4\oplus\hat{\dS}_4)$, where $\dS_4(\R)\simeq
I_{2,2}=\cl_{2,2}f$ is a minimal left ideal of the subalgebra
$\cl_{2,2}$, $f$ is a primitive idempotent of $\cl_{2,2}$. In
general, this spinor representation is realized within the matrix
algebra ${}^2\Mat_4(\R)$.

There is also a homomorphic mapping
\[
\epsilon:\;\cl_{3,2}\longrightarrow{}^\epsilon\cl_{2,2},
\]
where
\[
{}^\epsilon\cl_{2,2}\simeq\cl_{2,2}/\Ker\epsilon
\]
is a quotient algebra. In accordance with Theorem 14 in \cite{Var04}
at the mapping $\epsilon:\;\cl_{3,2}\rightarrow{}^\epsilon\cl_{2,2}$
we can transfer onto ${}^\epsilon\cl_{2,2}$ only the
antiautomorphism $\cA\rightarrow\widetilde{\cA}$ (time reversal
$T$). Therefore, we have
$\sAut({}^\epsilon\cl_{2,2})\simeq\{1,T\}\simeq\{\sI,\sE\}$. It is
easy to see that $\sAut({}^\epsilon\cl_{2,2})\simeq\{1,T\}$ is the
subgroup of $\sAut_-(\cl_{2,2})\simeq\dZ_4$ (see (\ref{Aut2_2}))
considered in the subsection 3.5. In the spinbasis (\ref{Basis5}),
defined by the idempotent $f^+_1\in\cl_{2,2}$, we obtain
\[
\sAut_-({}^\epsilon\cl_{2,2})\simeq\{1,T\}\simeq\{\boldsymbol{1}_4,\gamma_{34}\}\simeq\dZ_4/\dZ_2.
\]
The multiplication table of $\sAut_-({}^\epsilon\cl_{2,2})$ is given
in the Tab.\,7. The group $\dZ_4/\dZ_2\simeq\{1,i\}$ has element $i$ with order 4.
\begin{figure}[ht]
\[
{\renewcommand{\arraystretch}{1.4} \begin{tabular}{|c||c|c|}\hline
  & $1$ & $T$ \\ \hline\hline
$1$ & $1$ & $T$ \\ \hline $T$ & $T$ & $-1$ \\ \hline
\end{tabular}\;\;\sim\;\;
\begin{tabular}{|c||c|c|}\hline
  & $\boldsymbol{1}_4$ & $\gamma_{34}$ \\ \hline\hline
$\boldsymbol{1}_4$ & $\boldsymbol{1}_4$ & $\gamma_{34}$ \\
\hline
$\gamma_{34}$ & $\gamma_{34}$ & $-\boldsymbol{1}_4$ \\
\hline
\end{tabular}
}
\]
\hspace{0.3cm}
\begin{center}{\small \textbf{Tab.\,7:} The multiplication table of $\sAut_-({}^\epsilon\cl_{2,2})\simeq\dZ_4/\dZ_2$.}
\end{center}
\end{figure}

In conclusion it should be noted that anti-de Sitter space
$\R^{2,3}$ with the opposite signature $(+,+,-,-,-)$ corresponds to
the algebra $\cl_{2,3}$ of the type $p-q\equiv 7\pmod{8}$. This
algebra has the complex division ring $\K\simeq\C$ and, therefore,
there is an isomorphism $\cl_{2,3}\simeq\cl_4$. It is obvious that in
this case we return to $CPT$ groups considered in the section 3.

\section{Summary}
In this paper we have study discrete symmetries for the spinor field
in the spaces $\R^{4,1}$, $\R^{1,4}$ and $\R^{3,2}$ in terms of
automorphism groups of the algebras $\cl_{4,1}$, $\cl_{1,4}$ and
$\cl_{3,2}$. The algebras $\cl_{4,1}$, $\cl_{1,4}$ and $\cl_{3,2}$
are considered here as three model examples of the more general
algebraic framework. As is known, the spinor field (the field of the
spin-1/2) is described within $(1/2,0)\oplus(0,1/2)$-representation
of the Lorentz group. This representation is \emph{a fundamental
representation} of the group $\spin_+(1,3)$. For that reason Dirac
wrote that spin-1/2 is more elementary rather then spin-0
\cite{Dir}. All other (finite-dimensional) representations of
$\spin_+(1,3)$, which compound higher spin fields, are derived by
tensoring of spin-1/2 fields. There is a relationship between
representations of $\spin_+(1,3)$ and Clifford algebras
\cite{Var11,Var12}. At this point, complex representations of
$\spin_+(1,3)$ are described within a representation system
$\fM_\C=\fM^0\oplus\fM^1$, and real representations within a system
$\fM_\R=\fM^+\oplus\fM^-$, where

1) $\F=\C$, $\fM_\C=\fM^0\oplus\fM^1$.
\begin{eqnarray}
\fM^0&:&\fC^{l_0+l_1-1,l_0-l_1+1},\;\leftrightarrow \cl_n,\,n\equiv
0\pmod{2};\nonumber\\
\fM^1&:&{}^\epsilon\fC^{l_0+l_1-1,l_0-l_1+1},\;\leftrightarrow
\cl_n,\,n\equiv 1\pmod{2}.\nonumber
\end{eqnarray}

2) $\F=\R$, $\fM_\R=\fM^+\oplus\fM^-$.
\[
{\renewcommand{\arraystretch}{1.4} \fM^+:\;\left\{
\begin{array}{ccc}
\fR^{l_0}_0 & \leftrightarrow & \cl_{p,q},\;p-q\equiv
0\pmod{8},\;\K\simeq\R;\nonumber\\
\fR^{l_0}_2 & \leftrightarrow & \cl_{p,q},\;p-q\equiv
2\pmod{8},\;\K\simeq\R;\nonumber\\
\fH^{l_0}_4 & \leftrightarrow & \cl_{p,q},\;p-q\equiv
4\pmod{8},\;\K\simeq\BH;\nonumber\\
\fH^{l_0}_6 & \leftrightarrow & \cl_{p,q},\;p-q\equiv
6\pmod{8},\;\K\simeq\BH. \nonumber
\end{array}\right.
}
\]
\[
{\renewcommand{\arraystretch}{1.4} \fM^-:\;\left\{
\begin{array}{ccc}
\fC^{l_0}_3 & \leftrightarrow & \cl_{p,q},\;p-q\equiv
3\pmod{8},\;\K\simeq\C;\nonumber\\
\fC^{l_0}_7 & \leftrightarrow & \cl_{p,q},\;p-q\equiv
7\pmod{8},\;\K\simeq\C;\nonumber\\
\fR^{l_0}_{0,2}\oplus\fR^{l_0}_{0,2} & \leftrightarrow &
\cl_{p,q},\;p-q\equiv 1\pmod{8},\;\K\simeq\R\oplus\R;\nonumber\\
\fH^{l_0}_{4,6}\oplus\fH^{l_0}_{4,6} & \leftrightarrow &
\cl_{p,q},\;p-q\equiv 5\pmod{8},\;\K\simeq\BH\oplus\BH.\nonumber
\end{array}\right.
}
\]
Here the numbers $l_0$ and $l_1$ define a Gel'fand-Naimark
representation basis of the Lorentz group (for more details see
\cite{Var11,Var12}). Hence it follows that the system
$\fM_\C=\fM^0\oplus\fM^1$ and also the representations $\fC^{l_0}_3$
and $\fC^{l_0}_7$ of the block $\fM^-$ correspond to \emph{charged
particles}. The representations $\fH^{l_0}_4$ and $\fH^{l_0}_6$ of
the block $\fM^+$ and $\fH^{l_0}_{4,6}\oplus\fH^{l_0}_{4,6}$ of
$\fM^-$ (quaternionic representations of the group $\spin_+(1,3)$)
correspond to neutral particles of the first class (neutral
particles which admit \emph{particle-antiparticle interchange}). In
turn, the representations $\fR^{l_0}_0$ and $\fR^{l_0}_2$ of the
block $\fM^+$ and $\fR^{l_0}_{0,2}\oplus\fR^{l_0}_{0,2}$ of $\fM^-$
correspond to \emph{truly neutral particles}.

\section*{Appendix A: $CPT$ groups of the Dirac algebra $\cl_4$}
\setcounter{equation}{0}
\setcounter{section}{0}
\setcounter{subsection}{0}
\renewcommand{\thesubsection}{A.\arabic{subsection}}
\renewcommand{\theequation}{A.\arabic{equation}}

In this appendix we will define $CPT$ groups for
the algebra $\cl_{4,1}\simeq\cl_4$ in dependence on the division ring
structure of the real subalgebras $\cl_{p,q}\subset\cl_4$. We
consider in sequence $CPT$ groups generated by the subalgebras
$\cl_{1,3}$, $\cl_{4,0}$, $\cl_{0,4}$, $\cl_{3,1}$, $\cl_{2,2}$.
\subsection{The subalgebra $\cl_{1,3}$}
The primitive idempotent of the spacetime algebra $\cl_{1,3}$ can be
defined by an expression $f=\frac{1}{2}(1+\e_{14})$, and a division
ring has the form\footnote{Here we use the {\sc CLIFFORD} package
for Maple developed by R. Ab\l amowicz \cite{Abl98}.}
\[
\K=f\cl_{1,3}f=\{1,\,\e_2,\,\e_3,\,\e_{23}\}\simeq\{1,\,\bi,\,\bj,\,\bk\}\simeq\BH,
\]\begin{sloppypar}\noindent
where $\bi,\,\bj,\,\bk$ are well known quaternion units. Therefore,
the mapping
$\cl_{1,3}\overset{\gamma}{\longrightarrow}\End_{\BH}(\dS_2)$ leads
to the following representation:\end{sloppypar}
\begin{equation}\label{Basis1}
\gamma_1=\begin{bmatrix} 0 & \boldsymbol{1}_2\\
\boldsymbol{1}_2 & 0
\end{bmatrix},\;\;
\gamma_2=\begin{bmatrix} \e_2 & 0\\
0 & -\e_2
\end{bmatrix},\;\;
\gamma_3=\begin{bmatrix} \e_3 & 0\\
0 & -\e_3
\end{bmatrix},\;\;
\gamma_4=\begin{bmatrix} 0 & -\boldsymbol{1}_2\\
\boldsymbol{1}_2 & 0
\end{bmatrix}.
\end{equation}

Let us define $CPT$ group for the algebra $\cl_{1,3}$. First of all,
$\cl_{1,3}$ has the type $p-q\equiv 6\pmod{8}$, therefore, all the
eight automorphisms exist (see Theorem 10 in \cite{Var03}). Using
$\gamma$-matrices of the spinbasis (\ref{Basis1}), we define
elements of the group $\sExt(\cl_{1,3})$. The matrix of the
automorphism $\cA\rightarrow\cA^\star$ has the form
$\sW=\gamma_1\gamma_2\gamma_3\gamma_4=\gamma_{1234}$. Further, since
\[
\gamma^{t}_1=\gamma_1,\;\;\gamma^{t}_2=\gamma_2,\;\;\gamma^{t}_3=\gamma_3,\;\;
\gamma^{t}_4=-\gamma_4,
\]
then in accordance with $\widetilde{\sA}=\sE\sA^{t}\sE^{-1}$ we
have
\[
\gamma_1=\sE\gamma_1\sE^{-1},\;\;\gamma_2=\sE\gamma_2\sE^{-1},\;\;\gamma_3=\sE\gamma_3\sE^{-1},\;\;
\gamma_4=-\sE\gamma_4\sE^{-1} .
\]
Hence it follows that $\sE$ commutes with $\gamma_1$, $\gamma_2$,
$\gamma_3$ and anticommutes with $\gamma_4$, that is,
$\sE=\gamma_{123}$. From the definition $\sC=\sE\sW$ we find that
the matrix of the antiautomorphism
$\cA\rightarrow\widetilde{\cA^\star}$ has the form $\sC=\gamma_4$.
The spinbasis (\ref{Basis1}) contains both complex and real
matrices:
\[
\gamma^\ast_1=\gamma_1,\;\;\gamma^\ast_2=-\gamma_2,\;\;\gamma^\ast_3=-\gamma_3,\;\;
\gamma^\ast_4=\gamma_4,
\]
choosing $\e_2\sim\bi\sim\begin{bmatrix} 0 & -i\\
i & 0 \end{bmatrix}$, $\e_3\sim\bj\sim\begin{bmatrix} i & 0\\
0 & -i\end{bmatrix}$. Therefore, from
$\overline{\sA}=\Pi\sA^\ast\Pi^{-1}$ we obtain
\[
\gamma_1=\Pi\gamma_1\Pi^{-1},\;\;\gamma_2=-\Pi\gamma_2\Pi^{-1},\;\;\gamma_3=-\Pi\gamma_3\Pi^{-1},\;\;
\gamma_4=\Pi\gamma_4\Pi^{-1}.
\]
From the latter relations we find $\Pi=\gamma_{23}$. Further, in
accordance with $\sK=\Pi\sW$ for the spinor representation of the
pseudoautomorphism $\cA\rightarrow\overline{\cA^\star}$ we have
$\sK=\gamma_{14}$. Finally, for the pseudoantiautomorphisms
$\cA\rightarrow\overline{\widetilde{\cA}}$ ($CT$-transformation) and
$\cA\rightarrow\overline{\widetilde{\cA^\star}}$
($CPT$-transformation) from the definitions $\sS=\Pi\sE$ and
$\sF=\Pi\sC$ we find that $\sS=\gamma_1$ and $\sF=\gamma_{234}$.
Thus, we come to the following automorphism group:
\begin{multline}
\sExt(\cl_{1,3})\simeq\{\sI,\,\sW,\,\sE,\,\sC,\,\Pi,\,\sK,\,\sS,\,\sF\}\simeq\\
\simeq\{\boldsymbol{1}_4,\,\gamma_{1234},\,\gamma_{123},\,
\gamma_4,\,\gamma_{23},\,\gamma_{14},\,\gamma_1,\,\gamma_{234}\}.\nonumber
\end{multline}
The multiplication table of this group is given in the Tab.\,8. Super and subscripts in $\sExt^+_i$ correspond to super and subscripts of the primitive idempotents $f^+_i$.
\begin{figure}[ht]
\begin{center}{\renewcommand{\arraystretch}{1.4}
\begin{tabular}{|c||c|c|c|c|c|c|c|c|}\hline
  & $\boldsymbol{1}_4$ & $\gamma_{1234}$ & $\gamma_{123}$ & $\gamma_{4}$ & $\gamma_{23}$ &
$\gamma_{14}$ & $\gamma_{1}$ & $\gamma_{234}$\\ \hline\hline
$\boldsymbol{1}_4$ & $\boldsymbol{1}_4$ & $\gamma_{1234}$ &
$\gamma_{123}$ &
$\gamma_{4}$ & $\gamma_{23}$ & $\gamma_{14}$ & $\gamma_{1}$ & $\gamma_{234}$\\
\hline $\gamma_{1234}$ & $\gamma_{1234}$ & $-\boldsymbol{1}_4$ &
$\gamma_{4}$ &
$-\gamma_{123}$ & $-\gamma_{14}$ & $\gamma_{23}$ & $-\gamma_{234}$ & $\gamma_{1}$\\
\hline $\gamma_{123}$ & $\gamma_{123}$ & $-\gamma_{4}$ &
$-\boldsymbol{1}_4$ & $\gamma_{1234}$
& $-\gamma_{1}$ & $\gamma_{234}$ & $\gamma_{23}$ & $-\gamma_{14}$\\
\hline $\gamma_{4}$ & $\gamma_{4}$ & $\gamma_{123}$ &
$-\gamma_{1234}$ &
 $-\boldsymbol{1}_4$ & $\gamma_{234}$ & $\gamma_{1}$ & $-\gamma_{14}$ &
$-\gamma_{23}$\\ \hline $\gamma_{23}$ & $\gamma_{23}$ &
$-\gamma_{14}$ & $-\gamma_{1}$ & $\gamma_{234}$ &
$-\boldsymbol{1}_4$ & $\gamma_{1234}$ & $\gamma_{123}$ & $-\gamma_{4}$\\
\hline $\gamma_{14}$ & $\gamma_{14}$ & $\gamma_{23}$ &
$-\gamma_{234}$ & $-\gamma_{1}$
& $\gamma_{1234}$ & $\boldsymbol{1}_4$ & $-\gamma_{4}$ & $-\gamma_{123}$\\
\hline $\gamma_{1}$ & $\gamma_{1}$ & $\gamma_{234}$ & $\gamma_{23}$
& $\gamma_{14}$ & $\gamma_{123}$ & $\gamma_{4}$ & $\boldsymbol{1}_4$ & $\gamma_{1234}$\\
\hline $\gamma_{234}$ & $\gamma_{234}$ & $-\gamma_{1}$ &
$\gamma_{14}$ &
$-\gamma_{23}$ & $-\gamma_{4}$ & $\gamma_{123}$ & $-\gamma_{1234}$ & $\boldsymbol{1}_4$\\
\hline
\end{tabular}
}
\end{center}
\hspace{0.3cm}
\begin{center}
{\small \textbf{Tab.\,8:} The multiplication table of
$\sExt^+_1(\cl_{1,3})$.}
\end{center}
\end{figure}
As follows from this table the group $\sExt^+_1(\cl_{1,3})$ is an
non-Abelian finite group with the signature $(-,-,-,-,+,+,+)$. The
$CPT$ group
$C^{-,-,-,-,+,+,+}\simeq\overset{\ast}{\dZ}_4\times\dZ_2\times\dZ_2$
is the subgroup of $G(4,1)$. It is easy to verify that the same
group we obtain for the idempotent $f=\frac{1}{2}(1-\e_{14})$.

In common with the idempotents $f^\pm_1=\frac{1}{2}(1\pm\e_{14})$
the algebra $\cl_{1,3}$ has the following sequence of primitive
idempotents:
\[
f^\pm_2=\frac{1}{2}(1\pm\e_1),\;\;f^\pm_3=\frac{1}{2}(1\pm\e_{12}),\;\;
f^\pm_4=\frac{1}{2}(1\pm\e_{13}),\;\;f^\pm_5=\frac{1}{2}(1\pm\e_{234}).
\]
It is easy to verify that idempotents $f^+_2$, $f^+_3$, $f^+_4$ lead
with minor variations to the group
$\overset{\ast}{\dZ}_4\times\dZ_2\times\dZ_2$. We list below
division rings, spinbases and $CPT$ groups for $f^+_2$, $f^+_3$,
$f^+_4$:
\[
f^+_2=\frac{1}{2}(1+\e_1),
\]
\[
\K=f^+_2\cl_{1,3}f^+_2=\{1,\,\e_{23},\,\e_{24},\,\e_{34}\}\simeq\{1,\,\bi,\,\bj,\,\bk\}\simeq\BH,
\]
\[
\gamma_1=\begin{bmatrix} \boldsymbol{1}_2 & 0\\
0 & -\boldsymbol{1}_2
\end{bmatrix},\;
\gamma_2=\begin{bmatrix} 0 & -\boldsymbol{1}_2\\
\boldsymbol{1}_2 & 0
\end{bmatrix},\;
\gamma_3=\begin{bmatrix} 0 & \e_{23}\\
\e_{23} & 0
\end{bmatrix},\;
\gamma_4=\begin{bmatrix} 0 & \e_{24}\\
\e_{24} & 0
\end{bmatrix},
\]
\[
\sExt^+_2(\cl_{1,3})
\simeq\{\boldsymbol{1}_4,\,\gamma_{1234},\,\gamma_{134},\,
\gamma_2,\,\gamma_{34},\,\gamma_{12},\,\gamma_1,\,\gamma_{234}\}\simeq
\overset{\ast}{\dZ}_4\times\dZ_2,
\]
\[
C^{-,-,-,-,+,+,+}\simeq\overset{\ast}{\dZ}_4\times\dZ_2\times\dZ_2;
\]
\[
f^+_3=\frac{1}{2}(1+\e_{12}),
\]
\[
\K=f^+_3\cl_{1,3}f^+_3=\{1,\,\e_3,\,\e_4,\,\e_{34}\}\simeq\{1,\,\bi,\,\bj,\,\bk\}\simeq\BH,
\]
\[
\gamma_1=\begin{bmatrix} 0 & \boldsymbol{1}_2\\
\boldsymbol{1}_2 & 0
\end{bmatrix},\;\;
\gamma_2=\begin{bmatrix} 0 & -\boldsymbol{1}_2\\
\boldsymbol{1}_2 & 0
\end{bmatrix},\;\;
\gamma_3=\begin{bmatrix} \e_3 & 0\\
0 & -\e_3
\end{bmatrix},\;\;
\gamma_4=\begin{bmatrix} \e_4 & 0\\
0 & -\e_4
\end{bmatrix},
\]
\[
\sExt^+_3(\cl_{1,3})
\simeq\{\boldsymbol{1}_4,\,\gamma_{1234},\,\gamma_{134},\,
\gamma_2,\,\gamma_{34},\,\gamma_{12},\,\gamma_2,\,\gamma_{234}\}\simeq
\overset{\ast}{\dZ}_4\times\dZ_2,
\]
\[
C^{-,-,-,-,+,+,+}\simeq\overset{\ast}{\dZ}_4\times\dZ_2\times\dZ_2;
\]
\[
f^+_4=\frac{1}{2}(1+\e_{13}),
\]
\[
\K=f^+_4\cl_{1,3}f^+_4=\{1,\,\e_2,\,\e_4,\,\e_{24}\}\simeq\{1,\,\bi,\,\bj,\,\bk\}\simeq\BH,
\]
\[
\gamma_1=\begin{bmatrix} 0 & \boldsymbol{1}_2\\
\boldsymbol{1}_2 & 0
\end{bmatrix},\;\;
\gamma_2=\begin{bmatrix} \e_2 & 0\\
0 & -\e_2
\end{bmatrix},\;\;
\gamma_3=\begin{bmatrix} 0 & -\boldsymbol{1}_2\\
\boldsymbol{1}_2 & 0
\end{bmatrix},\;\;
\gamma_4=\begin{bmatrix} \e_4 & 0\\
0 & -\e_4
\end{bmatrix},
\]
\[
\sExt^+_4(\cl_{1,3})
\simeq\{\boldsymbol{1}_4,\,\gamma_{1234},\,\gamma_{124},\,
\gamma_3,\,\gamma_{24},\,\gamma_{13},\,\gamma_1,\,\gamma_{234}\}\simeq
\overset{\ast}{\dZ}_4\times\dZ_2,
\]
\[
C^{-,-,-,-,+,+,+}\simeq\overset{\ast}{\dZ}_4\times\dZ_2\times\dZ_2.
\]
It is not hard to see that idempotents $f^-_1$ $f^-_2$, $f^-_3$,
$f^-_4$ lead to the same group
$\overset{\ast}{\dZ}_4\times\dZ_2\times\dZ_2$.

In contrast to $f^\pm_1$ $f^\pm_2$, $f^\pm_3$, $f^\pm_4$, for the
primitive idempotents $f^\pm_5=\frac{1}{2}(1\pm\e_{234})$ we have
other realizations of $C^{a,b,c,d,e,f,g}$. Indeed,
\[
f^+_5=\frac{1}{2}(1+\e_{234}),
\]
\[
\K=f^+_5\cl_{1,3}f^+_5=\{1,\,\e_2,\,\e_3,\,\e_{23}\}\simeq\{1,\,\bi,\,\bj,\,\bk\}\simeq\BH,
\]
\begin{equation}\label{Basis2}
\gamma_1=\begin{bmatrix} 0 & \boldsymbol{1}_2\\
\boldsymbol{1}_2 & 0
\end{bmatrix},\;\;
\gamma_2=\begin{bmatrix} \e_2 & 0\\
0 & -\e_2
\end{bmatrix},\;\;
\gamma_3=\begin{bmatrix} \e_3 & 0\\
0 & -\e_3
\end{bmatrix},\;\;
\gamma_4=\begin{bmatrix} -\e_{23} & 0\\
0 & \e_{23}
\end{bmatrix}.
\end{equation}
In the spinbasis (\ref{Basis2}) we see that all the
$\gamma$-matrices are symmetric,
\[
\gamma^{t}_1=\gamma_1,\;\;\gamma^{t}_2=\gamma_2,\;\;\gamma^{t}_3=\gamma_3,\;\;
\gamma^{t}_4=\gamma_4.
\]
Therefore, the matrix $\sE$ of the antiautomorphism
$\cA\rightarrow\widetilde{\cA}$ commutes with all the matrices of
(\ref{Basis2}). This condition takes place only in the case when
$\sE\sim\boldsymbol{1}_4$. We suppose $\sE=\eta_t\boldsymbol{1}_4$,
where $\eta_t$ is a \emph{phase factor},
$\eta_t\in\C^\ast=\C-\{0\}$. Then from the definition $\sC=\sE\sW$
we have $\sC=\eta_t\gamma_{1234}$ for the antiautomorphism
$\cA\rightarrow\widetilde{\cA^\star}$. The basis (\ref{Basis2})
contains both complex and real matrices:
\[
\gamma^\ast_1=\gamma_1,\;\;\gamma^\ast_2=-\gamma_2,\;\;\gamma^\ast_3=-\gamma_3,\;\;
\gamma^\ast_4=-\gamma_4.
\]
Hence it follows that the matrix $\Pi$ of the pseudoautomorphism
$\cA\rightarrow\overline{\cA}$ commutes with $\gamma_1$ and
anticommutes with $\gamma_2$, $\gamma_3$, $\gamma_4$. Therefore, the
spinor representation of $\cA\rightarrow\overline{\cA}$ is defined
by $\Pi=\gamma_1$. Further, from the definition $\sK=\Pi\sW$ we have
$\sK=\gamma_{234}$ for the pseudoautomorphism
$\cA\rightarrow\overline{\cA^\star}$. Finally, from the definitions
$\sS=\Pi\sE$ and $\sF=\Pi\sC$ we obtain for the
pseudoantiautomorphisms $\cA\rightarrow\overline{\widetilde{\cA}}$
and $\cA\rightarrow\overline{\widetilde{\cA^\star}}$ the following
spinor representations: $\sS=\eta_t\gamma_1$ and
$\sF=\eta_t\gamma_{234}$. Thus, we come to the following generating
group:
\[
\sExt^+_5(\cl_{1,3})
\simeq\{\boldsymbol{1}_4,\,\gamma_{1234},\,\eta_t\boldsymbol{1}_4,\,
\eta_t\gamma_{1234},\,\gamma_1,\,\gamma_{234},\,\eta_t\gamma_1,\,
\eta_t\gamma_{234}\}.
\]
The multiplication table of this group is shown in the Tab.\,9.
\begin{figure}[ht]
{\footnotesize
\begin{center}{\renewcommand{\arraystretch}{1.4}
\begin{tabular}{|c||c|c|c|c|c|c|c|c|}\hline
  & $\boldsymbol{1}_4$ & $\boldsymbol{\omega}$ & $\eta_t\boldsymbol{1}_4$ & $\eta_t\boldsymbol{\omega}$ & $\gamma_{1}$ &
$\gamma_{234}$ & $\eta_t\gamma_{1}$ & $\eta_t\gamma_{234}$\\
\hline\hline $\boldsymbol{1}_4$ & $\boldsymbol{1}_4$ &
$\boldsymbol{\omega}$ & $\eta_t\boldsymbol{1}_4$ &
$\eta_t\boldsymbol{\omega}$ & $\gamma_{1}$ & $\gamma_{234}$ & $\eta_2\gamma_{1}$ & $\eta_t\gamma_{234}$\\
\hline $\boldsymbol{\omega}$ & $\boldsymbol{\omega}$ &
$-\boldsymbol{1}_4$ & $\eta_t\boldsymbol{\omega}$ &
$-\eta_t\boldsymbol{1}_4$ & $-\gamma_{234}$ & $\gamma_{1}$ & $-\eta_t\gamma_{234}$ & $\eta_t\gamma_{1}$\\
\hline $\eta_t\boldsymbol{1}_4$ & $\eta_t\boldsymbol{1}_4$ &
$\eta_t\boldsymbol{\omega}$ & $\eta^2_t\boldsymbol{1}_4$ &
$\eta^2_t\boldsymbol{\omega}$
& $\eta_t\gamma_{1}$ & $\eta_t\gamma_{234}$ & $\eta^2_t\gamma_{1}$ & $\eta^2_t\gamma_{234}$\\
\hline $\eta_t\boldsymbol{\omega}$ & $\eta_t\boldsymbol{\omega}$ &
$-\eta_t\boldsymbol{1}_4$ & $\eta^2_t\boldsymbol{\omega}$ &
 $-\eta^2_t\boldsymbol{1}_4$ & $-\eta_t\gamma_{234}$ & $\eta_t\gamma_{1}$ & $-\eta^2_t\gamma_{234}$ &
$\eta^2_t\gamma_{1}$\\ \hline $\gamma_{1}$ & $\gamma_{1}$ &
$\gamma_{234}$ & $\eta_t\gamma_{1}$ & $\eta_t\gamma_{234}$ &
$\boldsymbol{1}_4$ & $\boldsymbol{\omega}$ & $\eta_t\boldsymbol{1}_4$ & $\eta_t\boldsymbol{\omega}$\\
\hline $\gamma_{234}$ & $\gamma_{234}$ & $-\gamma_{1}$ &
$\eta_t\gamma_{234}$ & $-\eta_t\gamma_{1}$
& $-\boldsymbol{\omega}$ & $\boldsymbol{1}_4$ & $-\eta_t\boldsymbol{\omega}$ & $\eta_t\boldsymbol{1}_4$\\
\hline $\eta_t\gamma_{1}$ & $\eta_t\gamma_{1}$ &
$\eta_t\gamma_{234}$ & $\eta^2_t\gamma_{1}$ & $\eta^2_t\gamma_{234}$
& $\eta_t\boldsymbol{1}_4$ & $\eta_t\boldsymbol{\omega}$ &
$\eta^2_t\boldsymbol{1}_4$ &
$\eta^2_t\boldsymbol{\omega}$\\
\hline $\eta_t\gamma_{234}$ & $\eta_t\gamma_{234}$ &
$-\eta_t\gamma_{1}$ & $\eta^2_t\gamma_{234}$ & $-\eta^2_t\gamma_{1}$
& $-\eta_t\boldsymbol{\omega}$ & $\eta_t\boldsymbol{1}_4$ &
$-\eta^2_t\boldsymbol{\omega}$ &
$\eta^2_t\boldsymbol{1}_4$\\
\hline
\end{tabular}
}
\end{center}
} \hspace{0.3cm}
\begin{center}
\begin{minipage}{20pc}
{\small \textbf{Tab.\,9:} The multiplication table of
$\sExt^+_5(\cl_{1,3})$ in dependence of the sign of $\eta^2_t$,
where $\boldsymbol{\omega}=\gamma_{1234}$.
$\sExt^+_5(\cl_{1,3})\simeq D_4$ at $\eta^2_t=1$ and
$\sExt^+_5(\cl_{1,3})\simeq\overset{\ast}{\dZ}_4\times\dZ_2$ at
$\eta^2_t=-1$.}
\end{minipage}
\end{center}
\end{figure}
As follows from this table, when $\eta^2_t=1$ we have
$C^{-,+,-,+,+,+,+}\simeq D_4\times\dZ_2$ and
$C^{-,-,+,+,+,-,-}\simeq\overset{\ast}{\dZ}_4\times\dZ_2\times\dZ_2$
when $\eta^2_t=-1$. In the first case $\eta_t=\pm 1$, and in the
second case $\eta_t=\pm i$.
\subsection{The subalgebra $\cl_{4,0}$}
The next real subalgebra of $\cl_4$ is $\cl_{4,0}$. This algebra has
the type $p-q\equiv 4\pmod{8}$ and the first primitive idempotent
$f^+_1=\frac{1}{2}(1+\e_1)$. The division ring of $\cl_{4,0}$ is
\[
\K=f^+_1\cl_{4,0}f^+_1=\{1,\,\e_{23},\,\e_{24},\,\e_{34}\}\simeq\{1,\,\bi,\,\bj,\,\bk\}\simeq\BH,
\]
Therefore, the mapping
$\cl_{4,0}\overset{\gamma}{\longrightarrow}\End_{\BH}(\dS_2)$ gives
the following spinor representation:
\[
\gamma_1=\begin{bmatrix} \boldsymbol{1}_2 & 0\\
0 & -\boldsymbol{1}_2
\end{bmatrix},\;\;
\gamma_2=\begin{bmatrix} 0 & \boldsymbol{1}_2\\
\boldsymbol{1}_2 & 0
\end{bmatrix},\;\;
\gamma_3=\begin{bmatrix} 0 & -\e_{23}\\
\e_{23} & 0
\end{bmatrix},\;\;
\gamma_4=\begin{bmatrix} 0 & -\e_{24}\\
\e_{24} & 0
\end{bmatrix}.
\]
Making the same calculations as in the case of $\cl_{1,3}$, we come
to the group
\begin{equation}\label{Group^+_1}
\sExt^+_1(\cl_{4,0})\simeq\{\boldsymbol{1}_4,\,\gamma_{1234},\,\gamma_{34},\,
\gamma_{12},\,\eta_c\gamma_{34},\,\eta_c\gamma_{12},\,\eta_c\boldsymbol{1}_4,\,
\eta_c\gamma_{1234}\}.
\end{equation}
Hence it follows that
$\sExt^+_1(\cl_{4,0})\simeq\overset{\ast}{\dZ}_4\times\dZ_2$ at
$\eta^2_c=1$ and also
$\sExt^+_1(\cl_{4,0})\simeq\overset{\ast}{\dZ}_4\times\dZ_2$ at
$\eta^2_c=-1$. Therefore, in both cases we have
\begin{equation}\label{FGroup^+_1}
C^{+,-,-,-,-,+,+}\simeq
C^{+,-,-,+,+,-,-}\simeq\sExt^+_1(\cl_{4,0})\simeq\overset{\ast}{\dZ}_4\times\dZ_2\times\dZ_2.
\end{equation}
In common with the idempotent $f^+_1$ the algebra $\cl_{4,0}$ has
the following primitive idempotents:
\[
f^+_2=\frac{1}{2}(1+\e_2),\;\;f^+_3=\frac{1}{2}(1+\e_3),\;\;f^+_4=\frac{1}{2}(1+\e_4),\;\;
f^+_5=\frac{1}{2}(1+\e_{1234}).
\]
It is easy to verify that the idempotents $f^+_2$, $f^+_3$, $f^+_4$
lead to the group (\ref{Group^+_1}) (resp. (\ref{FGroup^+_1})), that
is, $\sExt^+_2(\cl_{4,0})$, $\sExt^+_3(\cl_{4,0})$,
$\sExt^+_4(\cl_{4,0})$ are isomorphic to $\sExt^+_1(\cl_{4,0})$.
However, situation is changed for the idempotent $f^+_5$. In this
case we have
\[
\K=f^+_5\cl_{4,0}f^+_5=\{1,\,\e_{23},\,\e_{24},\,\e_{34}\}\simeq\{1,\,\bi,\,\bj,\,\bk\}\simeq\BH,
\]
\[
\gamma_1=\begin{bmatrix} 0 & \boldsymbol{1}_2\\
\boldsymbol{1}_2 & 0
\end{bmatrix},\;\;
\gamma_2=\begin{bmatrix} 0 & \e_{34}\\
-\e_{34} & 0
\end{bmatrix},\;\;
\gamma_3=\begin{bmatrix} 0 & -\e_{24}\\
\e_{24} & 0
\end{bmatrix},\;\;
\gamma_4=\begin{bmatrix} 0 & \e_{23}\\
-\e_{23} & 0
\end{bmatrix},
\]
\begin{equation}\label{Group^+_5}
\sExt^+_5(\cl_{4,0})
\simeq\{\boldsymbol{1}_4,\,\gamma_{1234},\,\gamma_1,\,
\gamma_{234},\,\eta_c\gamma_1,\,\eta_c\gamma_{234},\,\eta_c\boldsymbol{1}_4,\,
\eta_c\gamma_{1234}\},
\end{equation}
and
\[
\eta^2_c=1:\;\sExt^+_5(\cl_{4,0})\simeq
D_4\,\longrightarrow\,C^{+,+,-,+,-,+,+}\simeq D_4\times\dZ_2,
\]
\[
\eta^2_c=-1:\;\sExt^+_5(\cl_{4,0})\simeq
\overset{\ast}{\dZ}_4\times\dZ_2\,\longrightarrow\,C^{+,+,-,-,+,-,-}\simeq
\overset{\ast}{\dZ}_4\times\dZ_2\times\dZ_2.
\]
It is not hard to see that for the idempotents $f^-_i$
$(i=1,\ldots,5)$ we obtain the same isomorphisms.

\subsection{The subalgebra $\cl_{0,4}$}
The next real subalgebra of $\cl_4$ with the quaternionic ring
$\K=\BH$ is $\cl_{0,4}$. This algebra has the type $p-q\equiv
4\pmod{8}$ and the first primitive idempotent
$f^+_1=\frac{1}{2}(1+\e_{123})$. The division ring of $\cl_{0,4}$
for $f^+_1$ is
\[
\K=f^+_1\cl_{0,4}f^+_1=\{1,\,\e_1,\,\e_{13},\,\e_3\}\simeq\{1,\,\bi,\,\bj,\,\bk\}\simeq\BH.
\]
Therefore, the mapping
$\cl_{0,4}\overset{\gamma}{\longrightarrow}\End_{\BH}(\dS_2)$ gives
the following spinor representation:
\begin{equation}\label{Basis3}
\gamma_1=\begin{bmatrix} \e_1 & 0\\
0 & -\e_1
\end{bmatrix},\;\;
\gamma_2=\begin{bmatrix} \e_{13} & 0\\
0 & -\e_{13}
\end{bmatrix},\;\;
\gamma_3=\begin{bmatrix} \e_3 & 0\\
0 & -\e_3
\end{bmatrix},\;\;
\gamma_4=\begin{bmatrix} 0 & -\boldsymbol{1}_2\\
\boldsymbol{1}_2 & 0
\end{bmatrix}.
\end{equation}
The generating group $\sExt^+_1(\cl_{0,4})$ arising from the
spinbasis (\ref{Basis3}) is
\[
\sExt^+_1(\cl_{0,4})
\simeq\{\boldsymbol{1}_4,\,\gamma_{1234},\,\gamma_{123},\,
\gamma_4,\,\eta_c\gamma_4,\,\eta_c\gamma_{123},\,
\eta_c\gamma_{1234},\,\eta_c\boldsymbol{1}_4\}.
\]
In this case we have
\[
\eta^2_c=1:\;\sExt^+_1(\cl_{0,4})\simeq
D_4\,\longrightarrow\,C^{+,+,-,-,+,+,+}\simeq D_4\times\dZ_2,
\]
\[
\eta^2_c=-1:\;\sExt^+_1(\cl_{0,4})\simeq
\overset{\ast}{\dZ}_4\times\dZ_2\,\longrightarrow\,C^{+,+,-,+,-,-,-}\simeq
\overset{\ast}{\dZ}_4\times\dZ_2\times\dZ_2.
\]
It is easy to verify that for other primitive idempotents
within $\cl_{0,4}$,
\[
f^\pm_2=\frac{1}{2}(1\pm\e_{124}),\;\;f^\pm_3=\frac{1}{2}(1\pm\e_{134}),\;\;f^\pm_4=\frac{1}{2}(1+\pm\e_{234}),
\]
\[
f^\pm_5=\frac{1}{2}(1\pm\e_{1234}),
\]\begin{sloppypar}\noindent
all the generating groups $\sExt^\pm_i(\cl_{0,4})$ $(i=2,\ldots,5)$
are isomorphic to $\sExt^+_1(\cl_{0,4})$.\end{sloppypar}

\subsection{The subalgebra $\cl_{3,1}$}
The first real subalgebra of $\cl_4$ with the real division ring
$\K=\R$ is $\cl_{3,1}$ (so-called \emph{Maiorana algebra}). This
algebra has the type $p-q\equiv 2\pmod{8}$ and the first
primitive idempotent is
\[
f^+_1=\frac{1}{4}(1+\e_1)(1+\e_{34})
\]
and
\[
\K=f^+_1\cl_{3,1}f^+_1\simeq\{1\}\simeq\R.
\]
The mapping
$\cl_{3,1}\overset{\gamma}{\longrightarrow}\End_{\R}(\dS_4)$ gives
\[
\gamma_1=\begin{bmatrix} 1 & 0 & 0 & 0\\
0 & -1& 0 & 0\\
0 & 0 & -1& 0\\
0 & 0 & 0 & 1
\end{bmatrix},\quad
\gamma_2=\begin{bmatrix} 0 & 1 & 0 & 0\\
1 & 0 & 0 & 0\\
0 & 0 & 0 & 1\\
0 & 0 & 1 & 0
\end{bmatrix},
\]
\begin{equation}\label{Basis4}
\gamma_3=\begin{bmatrix} 0 & 0 & 1 & 0\\
0 & 0 & 0 & -1\\
1 & 0 & 0 & 0\\
0 & -1& 0 & 0
\end{bmatrix},\quad
\gamma_4=\begin{bmatrix} 0 & 0 & -1& 0\\
0 & 0 & 0 & 1\\
1 & 0 & 0 & 0\\
0 & -1& 0 & 0
\end{bmatrix}.
\end{equation}
In case of the type $p-q\equiv 2\pmod{8}$ the matrix $\Pi$ of the
pseudoautomorphism $\cA\rightarrow\overline{\cA}$ is proportional to
the unit matrix (identical transformation) and the
automorphism group $\Ext(\cl_{p,q})$ is reduced to the group of
fundamental automorphisms, $\Aut_\pm(\cl_{p,q})$ (see Theorem 10 in
\cite{Var03}).

For the idempotent $f^+_1$ and the spinbasis (\ref{Basis4}) we have
\[
\sAut_+(\cl_{3,1})\simeq\{\boldsymbol{1}_4,\,\gamma_{1234},\,\gamma_{123},\,
\gamma_4\}\simeq Q_4/\dZ_2,
\]
\[
C^{-,-,-}\simeq Q_4.
\]
The idempotent $f^-_1=\frac{1}{4}(1-\e_1)(1-\e_{34})$ leads also to
the group $\sAut_+(\cl_{3,1})\simeq Q_4/\dZ_2$. Moreover, all other primitive idempotents of $\cl_{3,1}$,
\[
f^\pm_2=\frac{1}{4}(1\pm\e_1)(1\pm\e_{24}),\quad
f^\pm_3=\frac{1}{4}(1\pm\e_2)(1\pm\e_{14}),
\]
\[
f^\pm_4=\frac{1}{4}(1\pm\e_3)(1\pm\e_{134}),\quad
f^\pm_5=\frac{1}{4}(1\pm\e_3\e_4)(1\pm\e_{234})
\]
generate spinor representations with $\sAut_+(\cl_{3,1})\simeq
Q_4/\dZ_2$. Therefore, the group $C^{-,-,-}\simeq Q_4$ is an
invariant fact for $\cl_{3,1}$. In other words, this result does not
depend on the choice of the spinor representation.

\subsection{The subalgebra $\cl_{2,2}$}
Finally, we come to the subalgebra $\cl_{2,2}\subset\cl_4$. This
algebra has the type $p-q\equiv 0\pmod{8}$ and $\K\simeq\R$. The
mapping $\cl_{2,2}\overset{\gamma}{\longrightarrow}\End_{\R}(\dS_4)$
for the first primitive idempotent
$f^+_1=\frac{1}{4}(1+\e_{13})(1+\e_{24})$ gives
\[
\K=f^+_1\cl_{2,2}f^+_1\simeq\{1\}\simeq\R.
\]
\[
\gamma_1=\begin{bmatrix} 0 & 1 & 0 & 0\\
1 & 0& 0 & 0\\
0 & 0 & 0& 1\\
0 & 0 & 1 & 0
\end{bmatrix},\quad
\gamma_2=\begin{bmatrix} 0 & 0 & 1 & 0\\
0 & 0 & 0 & -1\\
1 & 0 & 0 & 0\\
0 & 1 & 0 & 0
\end{bmatrix},
\]
\begin{equation}\label{Basis5}
\gamma_3=\begin{bmatrix} 0 & -1 & 0 & 0\\
1 & 0 & 0 & 0\\
0 & 0 & 0 & -1\\
0 & 0& 1 & 0
\end{bmatrix},\quad
\gamma_4=\begin{bmatrix} 0 & 0 & -1& 0\\
0 & 0 & 0 & 1\\
1 & 0 & 0 & 0\\
0 & -1& 0 & 0
\end{bmatrix}.
\end{equation}
As in the previous case of $\cl_{3,1}$ the spinor representation of
the pseudoautomorphism $\cA\rightarrow\overline{\cA}$ is reduced to
the unit matrix. For the idempotents $f^\pm_1$ we have
\begin{equation}\label{Aut2_2}
\sAut_-(\cl_{2,2})\simeq\{\boldsymbol{1}_4,\,\gamma_{1234},\,\gamma_{34},\,\gamma_{12}\}
\simeq\dZ_4,
\end{equation}
\[
C^{+,-,-}\simeq\dZ_4\times\dZ_2.
\]
It is easy to verify that all other primitive idempotents of
$\cl_{2,2}$,
\[
f^\pm_2=\frac{1}{4}(1\pm\e_{23})(1\pm\e_{14}),\quad
f^\pm_3=\frac{1}{4}(1\pm\e_{14})(1\pm\e_{124}),
\]
\[
f^\pm_4=\frac{1}{4}(1\pm\e_{24})(1\pm\e_{1234}),\quad
f^\pm_5=\frac{1}{4}(1\pm\e_1)(1\pm\e_{23}),
\]
\[
f^\pm_6=\frac{1}{4}(1\pm\e_2)(1\pm\e_{13})
\]
generate spinor representations with
$\sAut_-(\cl_{2,2})\simeq\dZ_4$. Thus, the group
$C^{+,-,-}\simeq\dZ_4\times\dZ_2$ is an invariant fact for the
algebra $\cl_{2,2}$.

\section*{Acknowledgements}
I am grateful to Prof. R. Ab\l amowicz for helpful discussions about
{\sc CLIFFORD}.

\end{document}